\documentclass[sigconf]{acmart}

\AtBeginDocument{%
  }

\renewcommand\footnotetextcopyrightpermission[1]{} 

\usepackage{balance}
\usepackage{algorithmic}
\usepackage{graphicx}
\usepackage{textcomp}
\usepackage{xcolor}
\usepackage{xspace}
\usepackage{subcaption}
\usepackage{enumitem}
\usepackage{booktabs}
\usepackage{pifont}
\usepackage{siunitx}
\usepackage{multirow}

\usepackage[aboveskip=2pt, belowskip=-3pt]{caption}
\usepackage[ruled,vlined,linesnumbered,resetcount]{algorithm2e}

\newcommand{\sysname}{\textnormal{\textsc{REHSense}}\xspace}

\newcommand{\eg}{\textit{e.g.}\xspace}
\newcommand{\ie}{\textit{i.e.}\xspace}

\newcommand{\aka}{\textit{a.k.a.}\xspace}
\newcommand{\etal}{\textit{et al.}\xspace}

\newcounter{note}[section]

\colorlet{Mycolor1}{green!10!orange!90!}

\newcounter{packednmbr}

\renewcommand{\paragraph}[1]{\vspace{0.02in}\noindent{\bf{#1}}}
\begin{document}

\title{Towards Battery-Free Wireless Sensing via Radio-Frequency Energy Harvesting}

\author{Tao Ni$^{*}$, Zehua Sun$^{*}$, Mingda Han$^{\dag}$, Guohao Lan$^{\ddag}$, Yaxiong Xie$^{\P}$, Zhenjiang Li$^{*}$, Tao Gu$^{\parallel}$, Weitao Xu$^{*}$}
\affiliation{$^{*}$City University of Hong Kong, $^{\dag}$Shandong University, $^{\ddag}$Delft University of Technology, \\ $^{\P}$University at Buffalo, SUNY, $^{\parallel}$Macquarie University\country{}}
\begin{abstract}
Diverse Wi-Fi-based wireless applications have been proposed, 
ranging from daily activity recognition to vital sign monitoring. 
Despite their remarkable sensing accuracy, 
the high energy consumption and the requirement for customized hardware modification 
hinder the wide deployment of the existing sensing solutions.
In this paper, we propose \sysname, 
an energy-efficient wireless sensing solution based on Radio-Frequency (RF) energy harvesting. 
Instead of relying on a power-hungry Wi-Fi receiver, 
\sysname leverages an RF energy harvester as the sensor 
and utilizes the voltage signals harvested from the ambient Wi-Fi signals 
to enable simultaneous context sensing and energy harvesting. 
We design and implement \sysname using a commercial-off-the-shelf (COTS) RF energy harvester. 
Extensive evaluation of three fine-grained wireless sensing tasks (\ie, respiration monitoring, human activity, and hand gesture recognition) shows that 
\sysname can achieve comparable sensing accuracy with conventional Wi-Fi-based solutions 
while adapting to different sensing environments, reducing the power consumption by $98.7\%$ 
and harvesting up to \SI{4.5}{\milli\watt} of power from RF energy.
\end{abstract}

\maketitle
\pagestyle{plain} 

\vspace{0.1in}
\section{Introduction}
\label{sec:introduction}
Wi-Fi based wireless sensing systems are on the rise, 
with examples including health surveillance~\cite{liu2015contactless,zeng2018fullbreathe,wang2020csi}, activity recognition~\cite{zhang2019towards,wang2017device}, and gesture recognition~\cite{gao2022towards,he2015wig,niu2021understanding}. Among the existing systems, 
Channel State Information (CSI) based 
design has attracted great attention due to its promise in achieving fine-grained context sensing and 
compatibility with Commercial-Off-The-Shelf (COTS) Wi-Fi devices. 
However, in the existing Wi-Fi CSI-based sensing works\footnote{Complete list is available at: \url{https://github.com/REHSense/REHSense}},
we find that there are two practical limitations that hinder the wide adoption of Wi-Fi CSI-based sensing solutions. 
\looseness=-1

\begin{figure}[]
	\centering
	\begin{subfigure}[b]{.32\linewidth}
	    \centering
	    \includegraphics[width=\linewidth]{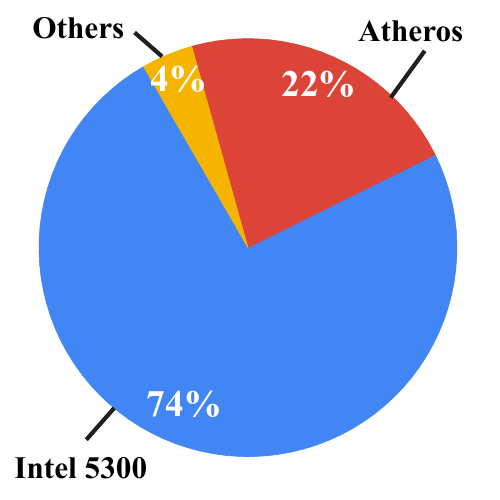}
	    \caption{Wi-Fi NICs.}
	    \label{fig:investigation-res1}
    \end{subfigure}
    \begin{subfigure}[b]{.32\linewidth}
        \centering
        \includegraphics[width=\linewidth]{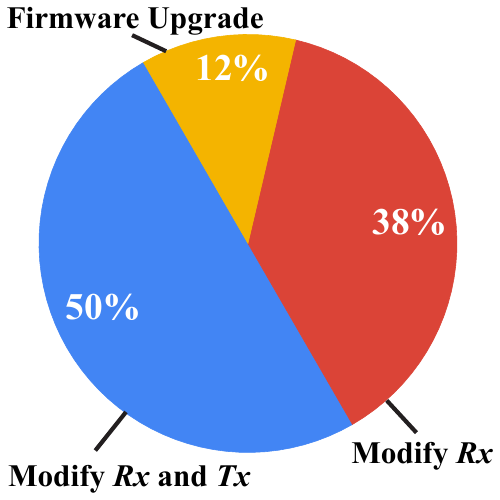}
        \caption{Modification.}
        \label{fig:investigation-res2}
    \end{subfigure}
    \begin{subfigure}[b]{.32\linewidth}
        \centering
        \includegraphics[width=\linewidth]{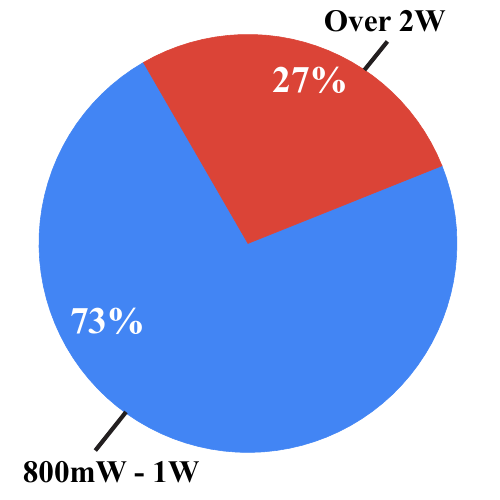}
        \caption{Consumption.}
        \label{fig:investigation-res3}
    \end{subfigure}
	\vspace{0.05in}
	\caption{\small Literature survey on $50$ representative CSI-based sensing works: (a) Wi-Fi NICs in investigated works; (b) Required hardware modifications; and (c) power consumption of $11$ Wi-Fi NICs.} 
	\label{fig:investigation-results}
	\vspace{-0.15in}
\end{figure}

\paragraph{(I) Customized hardware modification.} 
    First, 
    existing CSI-based sensing systems 
    utilize the Wi-Fi Network Interface Controller (NIC) card to receive the packets from the target Wi-Fi transmitter 
    and leverage the CSI tool~\cite{halperin2011tool,Xie:2015:PPD:2789168.2790124} to 
    extract the wireless channel information from the NIC. 
    However, not all NIC cards support CSI extraction. 
    In fact, most of the NIC manufacturers do not provide the access to CSI information. 
    As a result, we can only leverage a very limited range of NIC cards to obtain the CSI data for sensing, 
    i.e., the Intel $5300$ card~\cite{halperin2011tool} and the Qualcomm Atheros Wi-Fi chipset~\cite{Xie:2015:PPD:2789168.2790124}. 
    As shown in \autoref{fig:investigation-res1}, our literature survey indicates that $74\%$ of the 
    investigated CSI-based sensing works utilize the Intel $5300$ card~\cite{halperin2011tool}, 
    and $22\%$ of them use the Qualcomm Atheros Wi-Fi chipset~\cite{Xie:2015:PPD:2789168.2790124}.
    
    Moreover, since most of the CSI-extractable NIC cards are no longer adopted by the latest Wi-Fi devices, 
    existing solutions have to compromise the device. 
    This is done by replacing the original onboard NIC card in the device with a CSI-extractable alternate, 
    which is not only impractical but also requires additional effort from the end users. 
    As shown in \autoref{fig:investigation-res2}, $50\%$ of the existing 
    works need to modify both of the transceivers (Wi-Fi NICs in both router and mobile device), 
    $38\%$ of them need to modify the receiver, and the remaining $12\%$ of them require a firmware upgrade.
    
\paragraph{(II) High power consumption.} The second limitation of existing Wi-Fi CSI-based works is the high power consumption of the NICs. 
    Specifically, we investigate the power consumption of 11 CSI-extractable Wi-Fi NIC cards that are widely used by existing works~\footnote{Including the Intel 5300, three Intel AX200 Wi-Fi 6 series cards, and seven Qualcomm Atheros Wi-Fi chipsets.}. Their power consumption is briefly summarized in \autoref{fig:investigation-res3}. 
    Eight ($73\%$) of them consume approximately \SI{800}{\milli\watt} to \SI{1}{\watt} when running in the idle receiver mode~\cite{halperin2010demystifying,jang2011snooze}. 
    Three of them ($27\%$), i.e., the three Intel Wi-Fi NICs, consume over \SI{2}{\watt} to extract CSI information~
    \cite{jiang2021eliminating}. 
    Due to the high power consumption, 
    deploying Wi-Fi CSI-based sensing solutions on power-constrained devices, such as battery-powered IoT devices, becomes extremely difficult in a sustainable smart home.

To move beyond the current limitations and towards battery-free sensing, we present \sysname, a novel wireless sensing system based 
on \textit{\underline{\textbf{R}}F \underline{\textbf{E}}nergy \underline{\textbf{H}}arvesting}. 
Specifically, RF energy harvesting is a passive power scavenging technique that converts ambient RF energy emitted from mobile devices, 
such as Wi-Fi routers and cellular towers, into direct current (DC) voltage signals that can be used to power electrical components and devices. 
It has been widely used in diverse commodity IoT devices, such as LED lights~\cite{rehled}, tiny cameras~\cite{giordano2021battery}, and self-sustaining sensors~\cite{villani2021rf}.

\paragraph{Key insights.} Our key observation is that human activity significantly affects the amount of energy that the harvester harvests from the ambient RF signals,
since the signal reflecting off the human body may superimpose 
constructively or deconstructively with multipath signals in the propagation environment at the harvester, 
depending on the position and pose of the human.
Therefore, instead of purely utilizing it as a humdrum energy source, 
we propose to leverage the RF energy harvester as a novel battery-free activity sensor.  
Specifically, we measure the time-varying harvested power (voltage) as the sensing signal and   
apply a deep neural network to recognize human activities.  
We note that the relationship between the harvested energy and human activities has also been noticed in prior works~\cite{kellogg2014bringing, luo2019rf}.
However, we take the first step towards building battery-free Wi-Fi based sensing systems using purely low-cost off-the-shelf commercial hardware.

We design and implement \sysname by using low-cost commercial hardware, \textit{i.e.,} a Powercast P21XXCSR-EVB power harvesting board~\cite{ni2021simple} and an Arduino Nano microcontroller. 
We evaluate \sysname's performance with data collected from ten participants with four different COTS Wi-Fi routers in four typical environments (\ie, home, office, corridor, and cafe). 
Extensive experimental results show that \sysname achieves an accuracy of $95.4\%$, $95.7\%$, and $90.8\%$ 
in respiration monitoring, human activity recognition, and hand gesture recognition, respectively.
Furthermore, we conducted an experiment to compare the sensing accuracy and energy consumption of our energy-efficient solution 
with that of conventional sensing solutions using Wi-Fi NIC.
The results show that \sysname achieves a competitive sensing ability while consuming much lower energy than the CSI-based system. 
In addition, we present examples to demonstrate that \sysname can be integrated into different commodity IoT devices and provide a battery-free sensing prototype system based on RF energy harvesting.
\looseness=-1




\begin{figure}[]
    \centering
    \includegraphics[width=\linewidth]{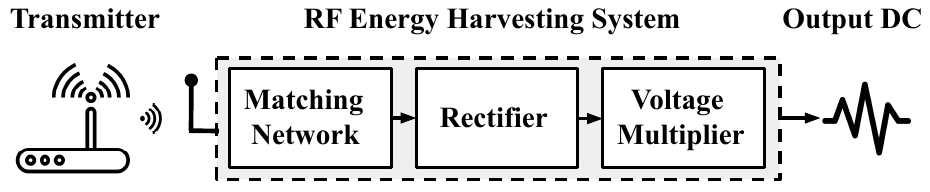}
    \vspace{0.05in}
    \caption{\small Workflow of RF energy harvesting.}
    \label{fig:rfeh-workflow}
    \vspace{-0.15in}
\end{figure}

\paragraph{Contributions.} We summarize the following contributions.

\begin{itemize}[leftmargin=10pt, itemsep=1pt, parsep=1pt]
    \item To the best of our knowledge, \sysname is the \textit{first} practical energy-efficient Wi-Fi sensing solution,
    which opens the door for designing battery-free wireless sensing systems.
    To provide a solid basis for future work, we also propose a theoretical model for describing RF energy transmission in the presence of various human activities in the Fresnel diffraction zones.
    \looseness=-1
    
   
    \item To support simultaneous energy harvesting and activity sensing, 
    we design a lightweight pipeline of \sysname that consists of the hardware platform, 
    signal processing modules, counting algorithms, and CNN-based neural networks. 
    \sysname reduces the energy consumption of wireless sensing systems by orders of magnitude while 
    imposing no hardware and firmware modifications to the existing Wi-Fi infrastructure. 
    
    \item 
    We design and implement a prototype of \sysname using COTS hardware circuits. 
    Our extensive experimental results demonstrate that \sysname achieves high accuracy in monitoring respiratory rate, 
    recognizing fine-grained human activities, and hand gestures. 
    Moreover, in a comparison experiment with a conventional CSI-based sensing system, 
    \sysname achieves competitive recognition accuracy in the same scenarios 
    while consuming significantly lower energy.
    \looseness=-1
\end{itemize}

\section{Preliminary}
\label{sec:background}

\subsection{Primer on RF Energy Harvesting}
\label{subsec:rfeh}


RF energy harvesting is a passive power scavenging technique that converts ambient RF energy emitted by high electromagnetic fields, such as Wi-Fi routers and cellular towers, into direct current (DC) that can be used to power sensors and 
prolong battery life. As an example, \autoref{fig:rfeh-workflow} shows a typical workflow of an RF energy harvesting system. The wireless transmitter (\eg, a Wi-Fi router) radiates electromagnetic energy to the near-field space~\cite{kim2014ambient}. 
The receiving antenna captures the RF signal and feeds it into an impedance-matching network that maximizes energy harvesting efficiency.
Then, a rectifier and a voltage multiplier are used to convert the harvested RF signal into DC voltage. 
Theoretically, the transmission of RF energy in free space follows the Friis' transmission equation~\cite{wadhwa2009impact} as follows: 
\begin{equation}
\label{eq:friis}
    P_{r} = P_{t}\frac{G_{r}G_{t}\lambda ^{2}}{(4\pi d)^{2}},
\end{equation}
where $P_{t}$, $P_{r}$ are the transmitted power and the received power; $G_{t}$, $G_{r}$ are the gains of the transmitting (Wi-Fi router) and the receiving (harvesting device) antennas; $\lambda$ is the wavelength; and $d$ is the distance between the two antennas. We can further obtain the gain of free-space RF energy transmission $Gain_{Friis}$ as: 
\begin{equation}
\label{eq:friisgain}
    Gain_{Friis}[dB] = 10log(\frac{P_{r}}{P_{t}}) = 10log(\frac{G_{t}G_{r}\lambda^2}{(4\pi d)^2}),
\end{equation}which describes the Line-of-Sight (LoS) harvested RF energy at distance $d$ between the transmitter and the receiver.
\looseness=-1

\subsection{Modeling Harvested Energy}
\label{subsec:fresnel}



The Fresnel zone diffraction model has been widely used in the analysis of fine-grained activity recognition~\cite{zhang2019towards,zhang2017toward}. The key insight is that when a target is located inside the Fresnel zones, diffraction dominates and becomes much stronger than other propagation effects (\eg, reflection and scattering). Prior studies~\cite{hristov2000fresnel}~\cite{zhang2021fresnel} have proved that more than $70\%$ of the RF energy is transmitted in the First Fresnel Zone (FFZ). Thus, we leverage the Fresnel zone diffraction model to study the RF energy harvesting voltage signal. 


\paragraph{Diffraction model formulation.} We first define the boundary of the $n$th Fresnel zone as: 
\begin{equation}
\label{eq:fresnelzone}
    \left | Tx P_{n} \right |+\left | Rx P_{n} \right |-\left | Tx Rx \right |=\frac{n\lambda }{2},
\end{equation}%
where the Wi-Fi router ($Tx$) and the RF energy harvesting device ($Rx$) are the foci of concentric ellipses and represent the Fresnel zones, $P_{n}$ is the sensing target located at the boundary of the $n$th Fresnel zone (\eg, $n=1$ represents the FFZ), and $\lambda$ is the wavelength of the RF signal.

\begin{figure}[]
    \centering
    \includegraphics[width=.9\linewidth]{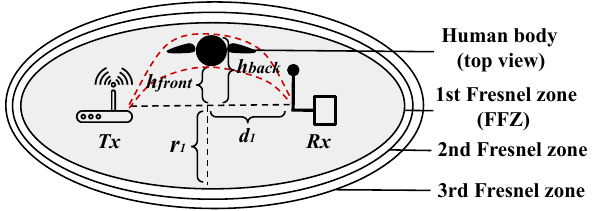}
    \caption{Illustration of the diffraction model  in the FFZ.}
    \vspace{-0.15in}
    \label{fig:fresnel-diffraction}
\end{figure}

\autoref{fig:fresnel-diffraction} illustrates the diffraction model in the FFZ. We assume the length of the semi-minor axis of the FFZ is $r_{1}$, and the sensing target is walking in the middle of $Tx$ and $Rx$ (i.e., $d_{1}=d/2$ and $r_{1}=\sqrt{\lambda d_{1}/2}$). The distance from the front and back sides of the human body to the LoS of the transceivers is $h_{front}$ and $h_{back}$, respectively. We can also obtain the {Fresnel front clearance} and the {Fresnel back clearance} as $u_{front}=h_{front}/r_{1}$  and $u_{back}=h_{back}/r_{1}$, respectively. Then, we obtain the Fresnel-Kirchhoff diffraction parameters $v_{front}=\sqrt{2}u_{front}$ and $v_{back}=\sqrt{2}u_{back}$~\cite{zhang2019towards}. Finally, the amplitude of the RF signals captured by the RF energy harvesting device, which is subject to the diffraction of the front and back sides of the human body, can be expressed as: 
\begin{equation}
\label{eq:diffamp}
\begin{split}
    F(v_{front}) = \frac{1+j}{2}\int_{v_{front}}^{\infty }e^{\frac{-j\pi z^{2}}{2}}dz, \\
    F(v_{back}) = \frac{1+j}{2}\int_{-\infty}^{v_{back}}e^{\frac{-j\pi z^{2}}{2}}dz,
\end{split}
\end{equation}%
where $e^{-j\pi z^2/2}$ is the phase shift in the diffraction path $z$.
When a human body is inside the FFZ, the total diffraction gain is the sum of diffraction gains of the front and back sides, which is defined as: 
\begin{equation}
\label{eq:diffgain}
    Gain_{Diff}\left [ dB \right ]=20log\left | F(v_{front})+F(v_{back}) \right |,
\end{equation}which describes the increases of harvested RF energy caused by the human body's diffraction effect.

\paragraph{Modeling the harvested energy.} So far, we have introduced the free-space gain of RF energy transmission $Gain_{Friis}$ and the gain of Fresnel diffraction $Gain_{Diff}$. In practice, however, there are many partitions and obstacles that can block the LoS RF energy transmission, and result in path losses. Thus, we further model the path loss at a distance $d$ as follows: 
\begin{equation}
\begin{split}
PL(d)\left[dB\right]&=PL(d_{0})+10nlog(\frac{d}{d_{0}}) +FAF\left[dB\right]\\
&+p\ast AF_{partition}\left[dB\right]+q\ast AF_{wall}\left[dB\right],
\label{eq:pathlossmodel}
\end{split}
\end{equation}%
where $PL(d_{0})$ is the path loss at reference distance (typically, $d_{0}=$ \SI{1}{\meter}), $FAF$ is the floor attenuation, $p$ and $q$ are the number of soft partitions (\ie, human body) and concrete walls, respectively, $AF_{partition}$ and $AF_{wall}$ are the corresponding attenuation factors, respectively. 
In addition, path loss describes the decrease of harvested RF energy when the human body totally or partially blocks LoS RF signal transmissions. 
\looseness=-1

In a nutshell, the energy generated by the RF energy harvesting device is the combination of the path losses $PL(d)$ and the two dominant gains, i.e., $Gain_{Friis}$ and $Gain_{Diff}$. Thus, leveraging \autoref{eq:friisgain}, \autoref{eq:diffgain} and \autoref{eq:pathlossmodel}, we can model the harvested energy and use it to achieve fine-grained activity recognition without the need of the conventional power-hungry wireless receiver. As shown later in our system profiling (\autoref{subsec:eval-comparison}), the proposed RF energy harvesting-based wireless sensing consumes only $11.3-$\SI{12.6}{\milli\watt} power, in contrast to the $820-$\SI{940}{\milli\watt} power consumed by conventional Wi-Fi NIC card. Below, we conduct a preliminary study to investigate the feasibility of the proposed idea.

\subsection{Feasibility Study}
\label{subsec:feasibility}

\begin{figure*}[]
	\centering
	\begin{subfigure}[b]{.325\linewidth}
	    \centering
	    \includegraphics[width=\linewidth]{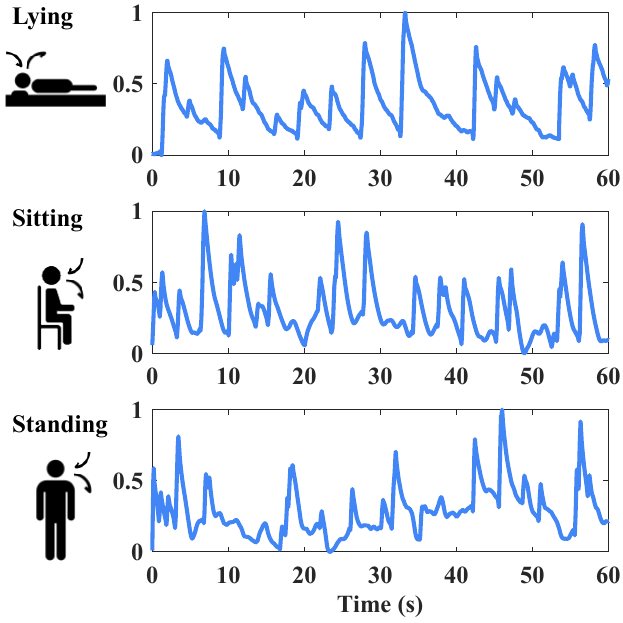}
	    \vspace{-0.2in}
	    \caption{Respiration monitoring.}
	    \label{fig:feasibility-study-rm}
    \end{subfigure}
    \begin{subfigure}[b]{.325\linewidth}
        \centering
        \includegraphics[width=\linewidth]{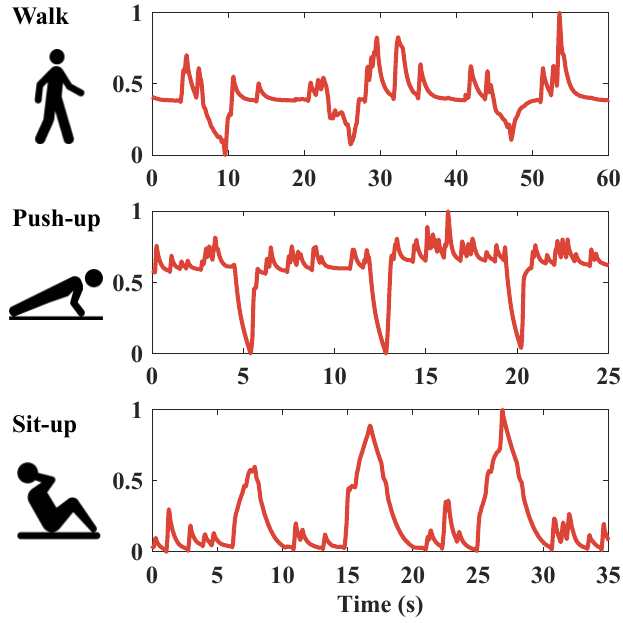}
        \vspace{-0.2in}
        \caption{Human activity recognition.}
        \label{fig:feasibility-study-har}
    \end{subfigure}
    \begin{subfigure}[b]{.325\linewidth}
        \centering
        \includegraphics[width=\linewidth]{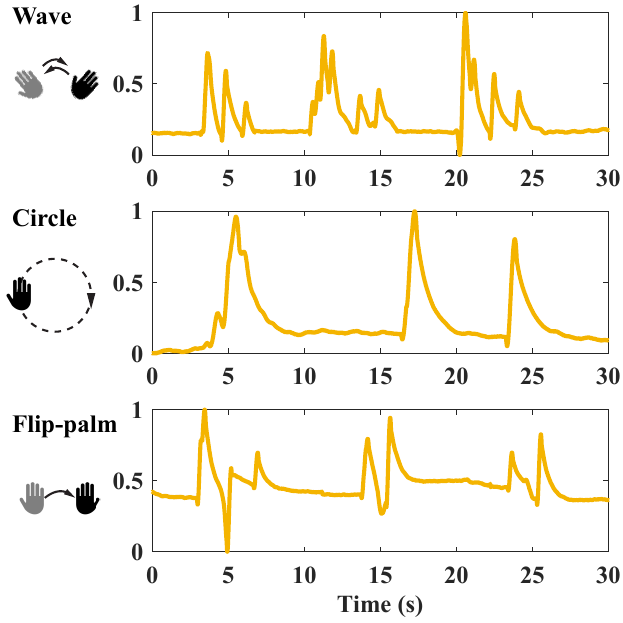}
        \vspace{-0.2in}
        \caption{Hand gesture recognition.}
        \label{fig:feasibility-study-hgr}
    \end{subfigure}
    \vspace{0.05in}
	\caption{Feasibility study: harvested voltage signals (normalized) in three wireless sensing tasks.}
	\label{fig:feasibility-study}
	\vspace{-0.1in}
\end{figure*}

We conduct a preliminary study to demonstrate the feasibility of using energy harvesting for wireless sensing. In this study, we consider three mainstream wireless sensing tasks: respiration monitoring, human activity recognition, and hand gesture recognition. For each sensing task, we monitor the voltage signals generated by an RF energy harvester when a subject is inside the FFZ between a commodity Wi-Fi router and an RF energy harvesting device (details of the hardware and setup are shown later in \autoref{fig:eval-experiment-setup}).


\begin{itemize}[leftmargin=10pt, itemsep=1pt, parsep=1pt]
\item \textbf{Case study 1: Respiration monitoring (RM).} The three plots in \autoref{fig:feasibility-study-rm} are the normalized voltage signals when a participant is breathing normally in different positions, including lying, sitting, and standing. The harvested signals show the continuous rise and fall patterns that correspond to the subtle displacements of the subject's chest. Recall our modeling in the previous subsection, the subtle displacement leads to changes in the diffraction gain in the FFZ, as well as the harvested RF energy.

\item \textbf{Case study 2: Human activity recognition (HAR).} \autoref{fig:feasibility-study-har} shows the harvested voltage signals 
when the person is performing three daily activities: walk, push-up, and sit-up. The signal patterns can also be illustrated by the diffraction model in \autoref{subsec:fresnel}. Taking walking as an example, when the person is approaching the FFZ, the diffraction gain increases, and hence more RF energy is captured by the receiver. Then, the person arrives in the middle of the FFZ, and the human body blocks the LoS energy transmission which increases the energy of path loss. Finally, as the person moves out of the FFZ, the harvested RF energy increases because the energy caused by path loss increases and the diffraction effect strengthens the harvested RF energy. Using the same principle, we can explain the patterns exhibited in the harvested signals of push-up and sit-up.

\item \textbf{Case study 3: Hand gesture recognition (HGR).} \autoref{fig:feasibility-study-hgr} shows the harvested RF energy signals of three hand gestures: waving hand, drawing a circle, and flipping the palm. We observe that different hand gestures lead to different bursts of captured RF energy because the moving hand changes the diffraction gain in the FFZ. Therefore, it is feasible to recognize fine-grained hand gestures using RF energy harvesting.
\end{itemize}

The preliminary results from the three case studies above show that changes in harvested RF energy can reflect human activities and vital signals such as respiratory rate. Moreover, different human activities (\ie, walking, push-ups, sit-ups) and human hand gestures inside the FFZ also present distinct RF energy patterns, demonstrating the feasibility of harnessing RF energy harvesting-based wireless sensing.
\looseness=-1

\section{Design and Implementation}
\label{sec:design-implementation}

\subsection{System Overview}
\label{subsec:overview}

Based on the proposed diffraction model and the results of the feasibility study, 
we present \sysname, a novel wireless sensing system via RF energy harvesting. 
\sysname implements a lightweight sensing pipeline that supports diverse sensing applications. 
We choose to implement three sensing tasks listed in our feasibility study for two reasons. 
First, they are the most representative applications in the wireless human sensing field~\cite{liu2019wireless}. 
Second, they cover different levels of granularity: respiration monitoring (minor movement), 
human activity recognition (large-scale movement), and hand gesture recognition (small-scale movement). 
Therefore, these three sensing tasks can demonstrate the potential of \sysname in enabling various applications for wireless sensing tasks.
\looseness=-1



\begin{figure}[]
    \centering
    \includegraphics[width=\linewidth]{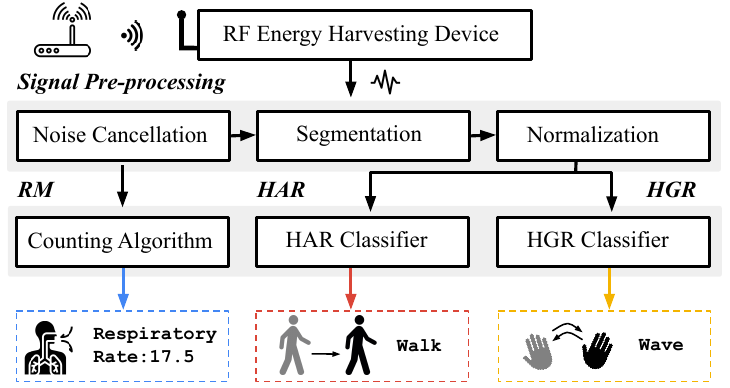}
    \vspace{-0.1in}
    \caption{Overview of \sysname.}
    \label{fig:system-overview}
    \vspace{-0.15in}
\end{figure}

\autoref{fig:system-overview} depicts the system overview of \sysname, 
which consists of three components. 
First, the RF energy harvesting device captures the RF signals sent by the Wi-Fi router 
and converts the RF signals to DC output (time-series voltages). 
Second, the signal pre-processing module performs noise cancellation to remove interference, segmentation to obtain segments containing activities and gestures, 
and normalization to generate valid input for training or recognition. 
Then, the counting algorithm takes the filtered voltage signal and outputs the calculated respiratory rate to achieve respiration monitoring (RM). 
The pre-trained CNN-based HAR and HGR classifiers can predict the activities and gestures from the normalized signal segments, respectively.
\looseness=-1

\begin{figure}[ht]
    \centering
\end{figure}




\subsection{Signal Pre-processing}
\label{subsec:hardware-signalpreprocessing}


\paragraph{Noise cancellation.} 
The harvested voltage signal contains noise that results from other transmitter sources and ambient interference. In \sysname, we apply a Savitzky-Golay (S-G) filter~\cite{luo2005savitzky} for noise filtering.
S-G filter fits successive slices of signal with a low degree polynomial function to maintain the time and frequency domain patterns of the time-series signal~\cite{acharya2016application}. Thus, it can effectively remove both high-frequency noise and small variations in the harvested signal caused by interference from other transmitters.

\paragraph{Signal segmentation.} 
We utilize a variance-based sliding window (\ie, \SI{0.5}{\second} with a $50\%$ overlapping rate) to divide the time-series voltage signal into segments. Specifically, the variance-based sliding window finds the segmentation points where the variance is lower than a pre-defined threshold (\ie, $0.1$ in our current design). The threshold is selected when no activity movements are performed, and the corresponding harvested voltage signal remains constant.
\looseness=-1

\paragraph{Normalization.} According to \autoref{eq:friis}, the distance between the two antennas affects the strength of the harvested RF energy. To reduce its impact on activity and gesture recognition, we normalize the signal segments to the range of $[0,1]$ by deducting the DC offset (average voltage where variance is less than $0.01$). In addition, the length of activity/gesture segments is different, so we apply the down-sampling method to resize the segments with different lengths to $1\times 128$ vectors as valid input of the CNN neural network in both training and testing processes.

\subsection{Respiration Monitoring (RM)}
\label{subsec:implementation-respiration}

In \autoref{subsec:feasibility}, we present that the RF signals of human respiration exhibit periodic motions that correspond to the rise and fall of the chest. However, we also find that some voltage spikes overlap together as the respiratory rate can be fast, and hence the harvested RF energy is a combination of the neighbor respiration. To address this overlap issue and achieve accurate respiration monitoring, we propose a \textit{variance-peak} counting algorithm (see \autoref{alg:variance-peak-counting}) that can calculate the respiratory rate by a moving-variance window and peak analysis. Specifically, it first obtains the variance signal by applying a moving-variance sliding window (Line $2-5$) and then finds peaks in the variance signal and uses the number of peaks to calculate the final respiratory rate (Line $6-7$). In \sysname, we set the size of the moving-variance window as $100$ with a given variance threshold as $0.002$, and use the peak analysis function provided by the Signal Processing Toolbox of MATLAB (R2022a).

\begin{algorithm}[t]
\caption{Variance-Peak Counting Algorithm}
\label{alg:variance-peak-counting}
\DontPrintSemicolon

\KwIn{$\mathcal{S}$: filtered harvested voltage signal. \\ \quad \quad \quad $w$: moving-variance window size. \\ \quad \quad \quad $\tau$: given variance threshold.}

\KwOut{$r$: calculated respiratory rate.}

$\mathcal{V}\leftarrow \mathcal{V}_{0}$, $\mathcal{V}_{0}=\phi $ \algorithmiccomment{initialize an empty variance array}

\For{$i=0; i<len(\mathcal{S})-w+1; i=i+1$}{

$\mathcal{V}_{i}= variance(\mathcal{S}_{i}, \mathcal{S}_{i+w})$ \algorithmiccomment{variance of a window}

\If{$\mathcal{V}_{i}\geq \tau$}{
$\mathcal{V}\leftarrow \mathcal{V}_{i}$ \algorithmiccomment{append $\mathcal{V}_{i}$ to $\mathcal{V}$}
}
}

$\mathcal{P} = findpeaks(\mathcal{V})$  \algorithmiccomment{find peaks in variance signal}

Output respiratory rate $r$ with $r=N_{\mathcal{P}}/t_{\mathcal{S}}$ \algorithmiccomment{$N_{\mathcal{P}}$ is the number of peaks and $t_{\mathcal{S}}$ is the time duration.}
\end{algorithm}

\begin{figure}[]
    \centering
    \includegraphics[width=\linewidth]{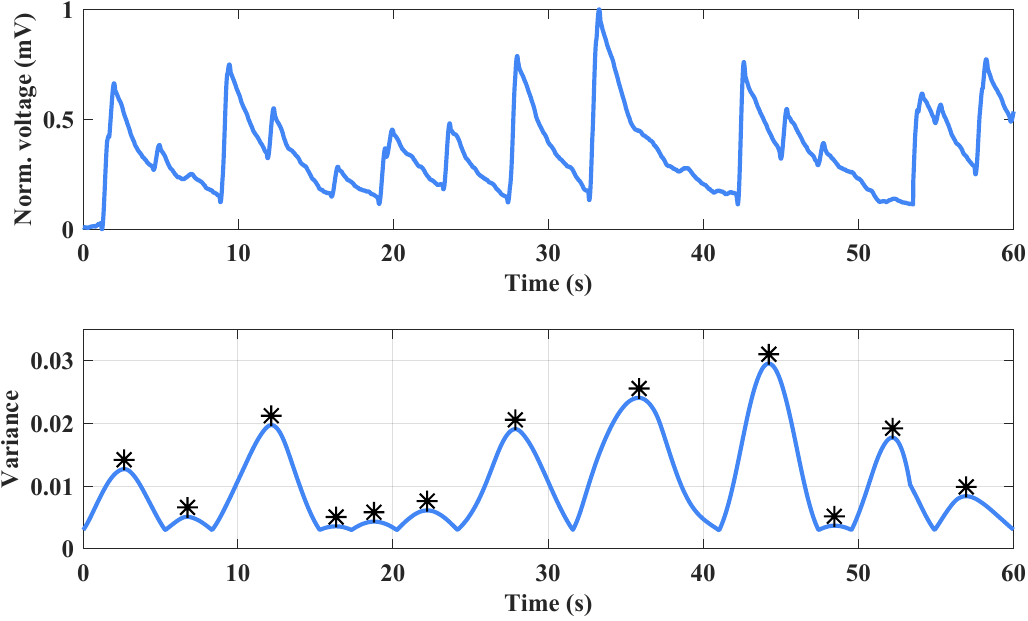}
    \caption{An example showing the performance of the \textit{variance-peak} respiration counting algorithm. The number of black asterisks denotes the number of respiration during the monitoring period.}
    \label{fig:example-variance-peak-counting}
    \vspace{-0.1in}
\end{figure}

\autoref{fig:example-variance-peak-counting} presents an example of applying the \textit{variance-peak} counting algorithm to calculate the respiration rate when a person is lying down on a mat. The upper figure shows the original harvested voltage signal and the lower figure presents the results after applying the proposed algorithm. It can be seen that compared to the original noisy signal in the upper figure, the peaks which correspond to breathing rate can be easily spotted in the lower figure after using our algorithm.

\subsection{Human Activity and Hand Gesture Recognition (HAR and HGR)}
\label{subsec:implementation-har-hgr}

\begin{figure}[]
    \centering
    \includegraphics[width=\linewidth]{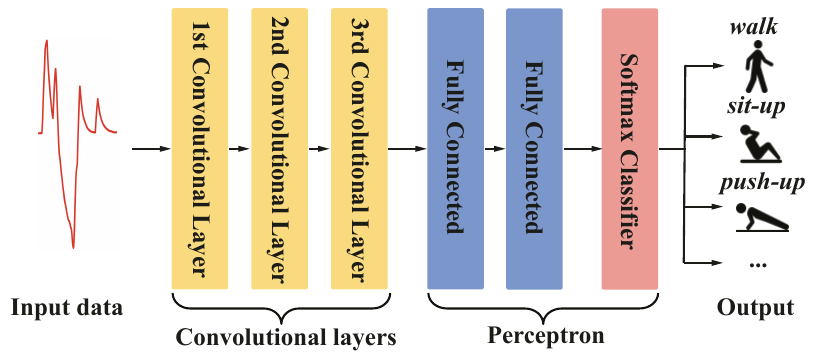}
    \caption{Architecture of the CNN neural network.}
    \vspace{-0.15in}
    \label{fig:cnn-model}
\end{figure}

To determine the type of human activity or hand gesture, we apply a Convolutional Neural Network (CNN) method to achieve activity and gesture classification in \sysname. Because CNN-based neural network supports classic classification models that can extract both temporal and spatial features from the harvested voltage signals while achieving high accuracy in class prediction. \autoref{fig:cnn-model} shows the architecture of the proposed CNN neural network that consists of six layers: three convolutional layers for extracting temporal and spatial features from the one-dimensional signal segments, two fully connected layers that take the extracted feature vectors to generalize the neural network and learn the non-linear combinations of these features. At last, a softmax function is utilized in the classifier to output the predicted activity/gesture type.

Specifically, each convolution layer uses ReLU as the activation function and is followed by a max pooling layer with a pool size of $2$. A flatten layer is used to convert the feature map to a one-dimensional feature vector, fed into the fully-connected layer. A dropout layer is used after the first fully-connected layer to avoid over-fitting, and the dropout rate is set to $0.5$. We implement the proposed CNN model in Keras 2.3 on the Tensorflow 2.0 framework and train $300$ epochs in total with an initial learning rate of $0.01$.
\looseness=-1

\section{Evaluation}
\label{sec:evaluation}

\subsection{Experimental Setup}
\label{subsec:eval-experiment}

\begin{figure}[]
    \centering
	\begin{subfigure}[b]{.49\linewidth}
        \centering
        \includegraphics[width=\linewidth]{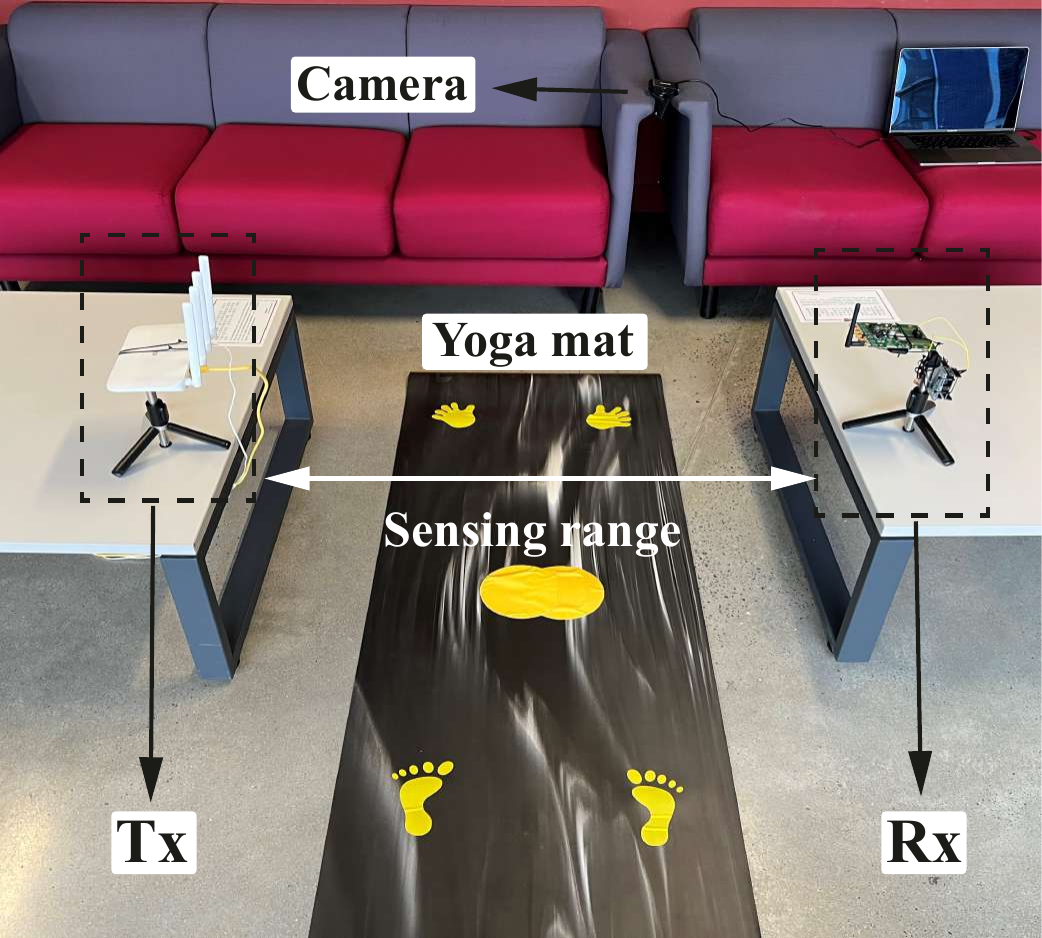}
        \caption{Experimental setup.}
        \label{fig:experiment-scenario}
    \end{subfigure}
	\begin{subfigure}[b]{.49\linewidth}
	    \centering
	    \includegraphics[width=\linewidth]{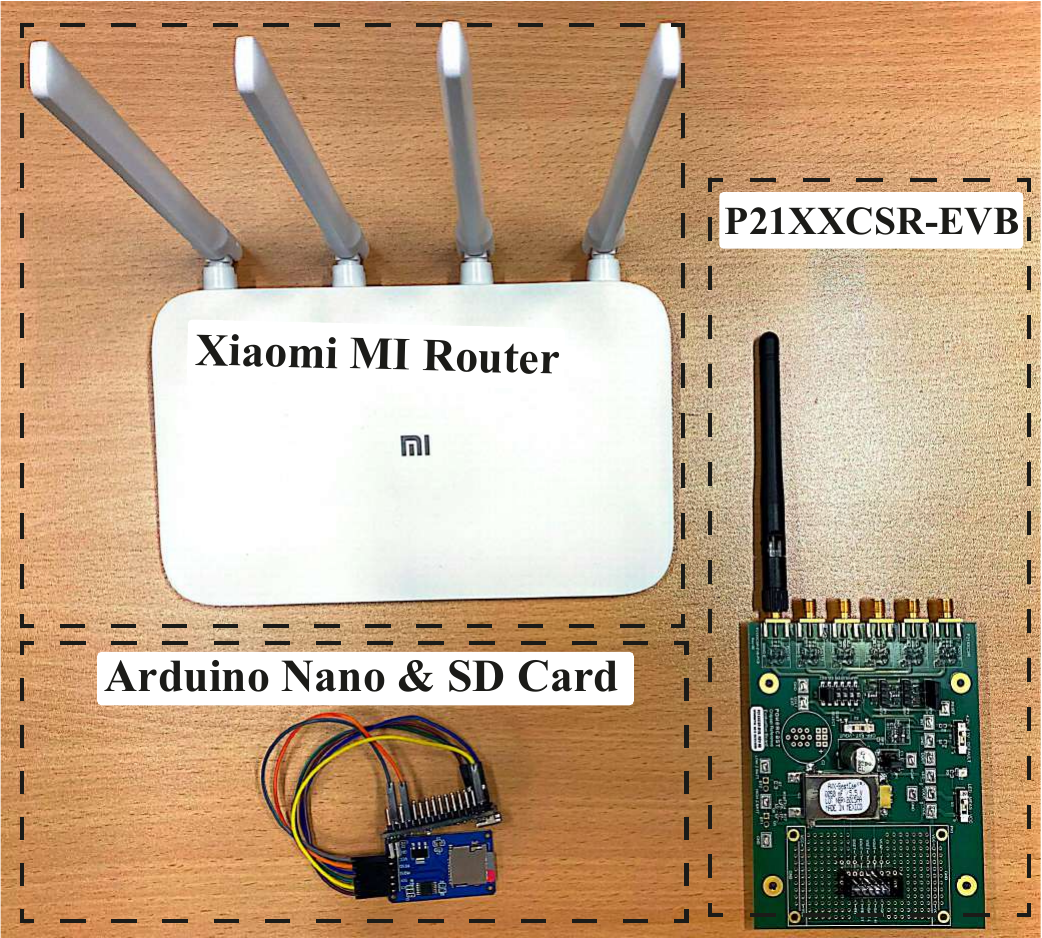}
	    \caption{Hardware devices.}
	    \label{fig:experiment-hardware}
    \end{subfigure}
	\vspace{0.05in}
	\caption{Experimental setup and hardware devices.}
        \vspace{-0.2in}
	\label{fig:eval-experiment-setup}
\end{figure}

\autoref{fig:eval-experiment-setup} shows the experimental environment, setup, and the hardware devices used in the experiments. We use a Xiaomi MI Router as the transmitter ($Tx$) and implement a prototype of RF energy harvesting as the receiver ($Rx$). In the prototype, we use an RF energy harvesting evaluation board, Powercast P21XXCSR-EVB~\cite{ni2021simple}, with one \SI{2.4}{\giga\hertz}, and one \SI{2.3}{\deci\bel}i dipole antennas to harvest RF energy from. To record the harvested voltage data, we use an Arduino Nano microcontroller and a \SI{32}{\giga\byte} MicroSD card as the data recorder. The sampling rate of the data recorder is \SI{200}{\hertz}. We consider the experiments in an indoor scenario (\eg, room \SI{5.7}{\meter}$\times$\SI{4.2}{\meter}) and set the sensing range between $Tx$ and $Rx$ to \SI{1.0}{\meter} and the height of $Tx$ and $Rx$ to \SI{0.5}{\meter}. We also place a camera to record the experimental process, which will be used as a reference to analyze the results.
All data processing and model training are conducted on a desktop running Windows $10$ with \SI{32}{\giga\byte} memory and an Intel i7-9700K CPU and an NVIDIA GeForce RTX 2080Ti GPU. In addition, we conduct experiments to explore the performance of \sysname in different settings (\ie Wi-Fi routers, sensing ranges, environments) in \autoref{subsec:impact-factors}.
\looseness=-1

\begin{figure*}[]
	\centering
	\begin{subfigure}[b]{.325\linewidth}
	    \centering
	    \includegraphics[width=\linewidth]{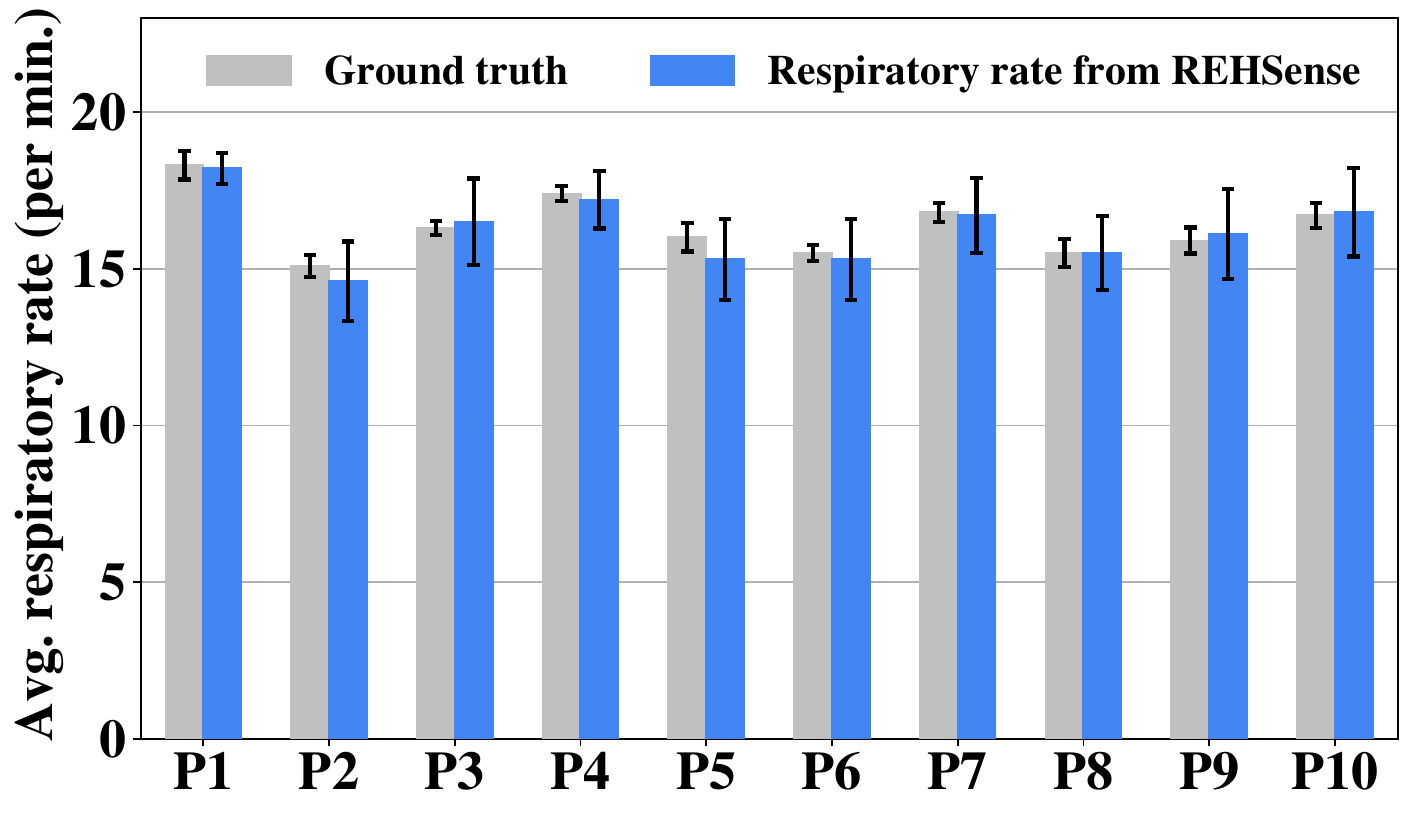}
	    \vspace{-0.2in}
	    \caption{Lying.}
	    \label{fig:effectiveness-rm-lying}
    \end{subfigure}
    \begin{subfigure}[b]{.325\linewidth}
        \centering
        \includegraphics[width=\linewidth]{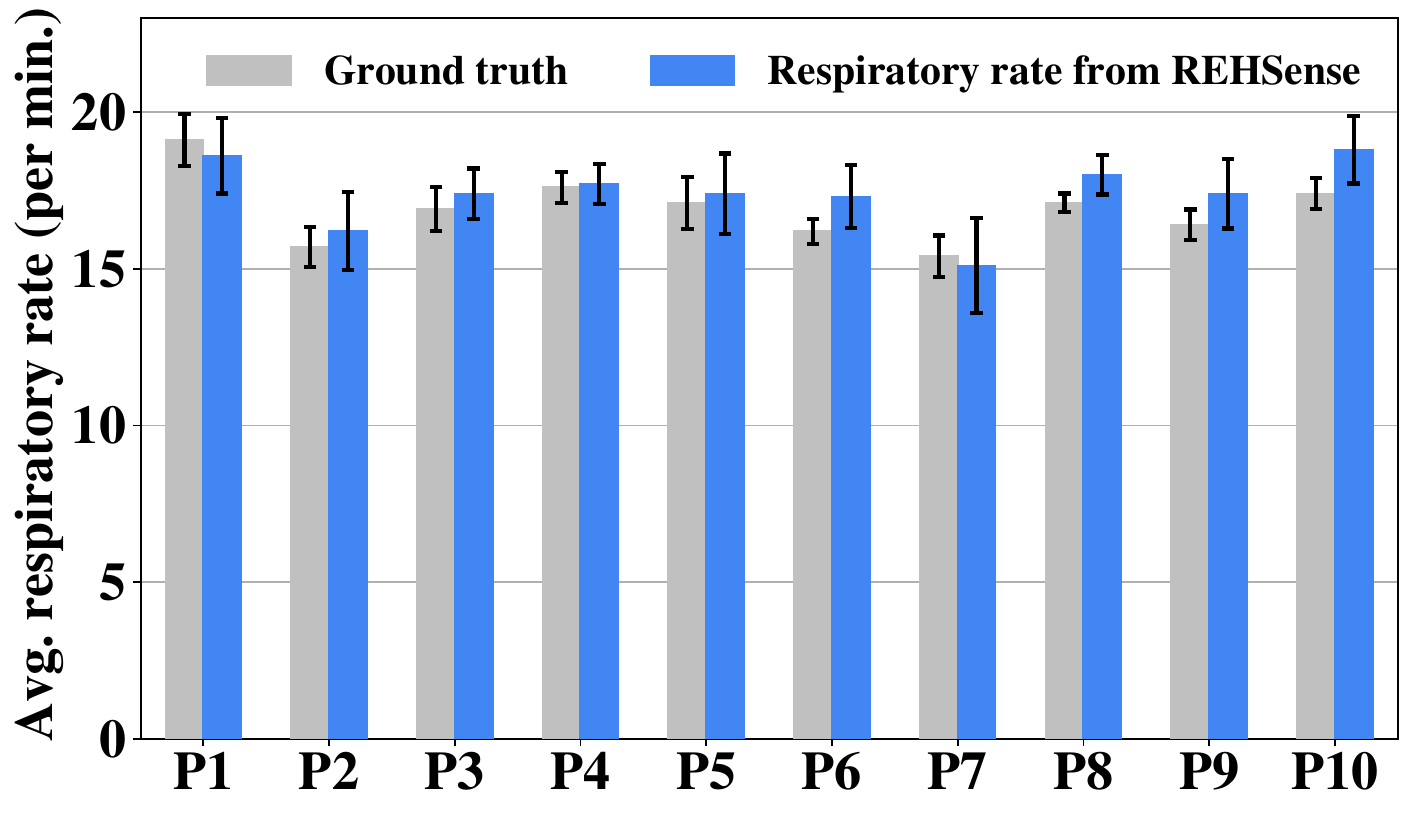}
        \vspace{-0.2in}
        \caption{Sitting.}
        \label{fig:effectiveness-rm-sitting}
    \end{subfigure}
    \begin{subfigure}[b]{.325\linewidth}
        \centering
        \includegraphics[width=\linewidth]{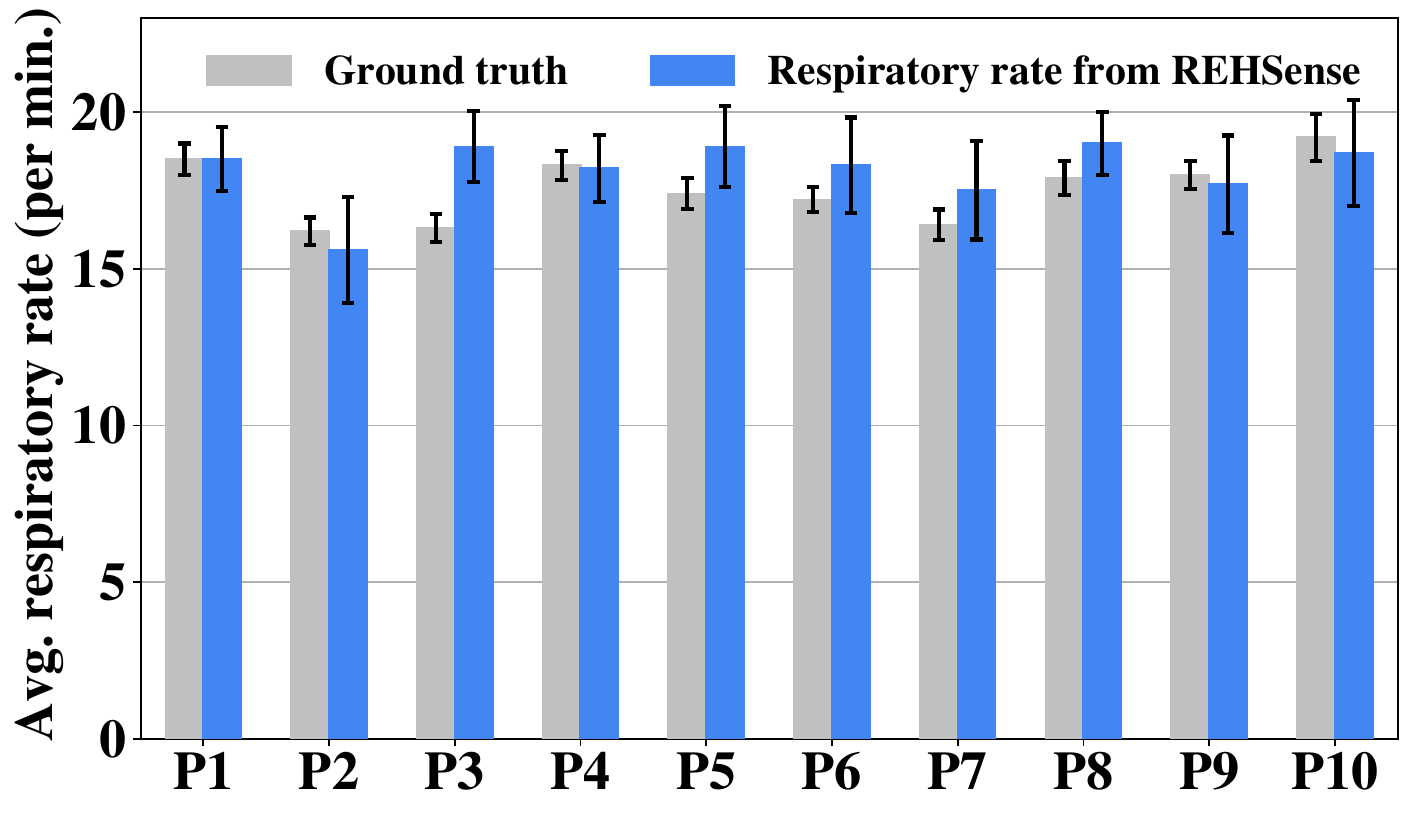}
        \vspace{-0.2in}
        \caption{Standing.}
        \label{fig:effectiveness-rm-standing}
    \end{subfigure}
	\vspace{0.05in}
	\caption{Effectiveness of respiration monitoring in three situations.}
        \vspace{-0.1in}
	\label{fig:effectiveness-rm-result}
\end{figure*}


\begin{figure*}[]
	\centering
	\begin{subfigure}[b]{.49\linewidth}
	    \centering
	    \includegraphics[width=\linewidth]{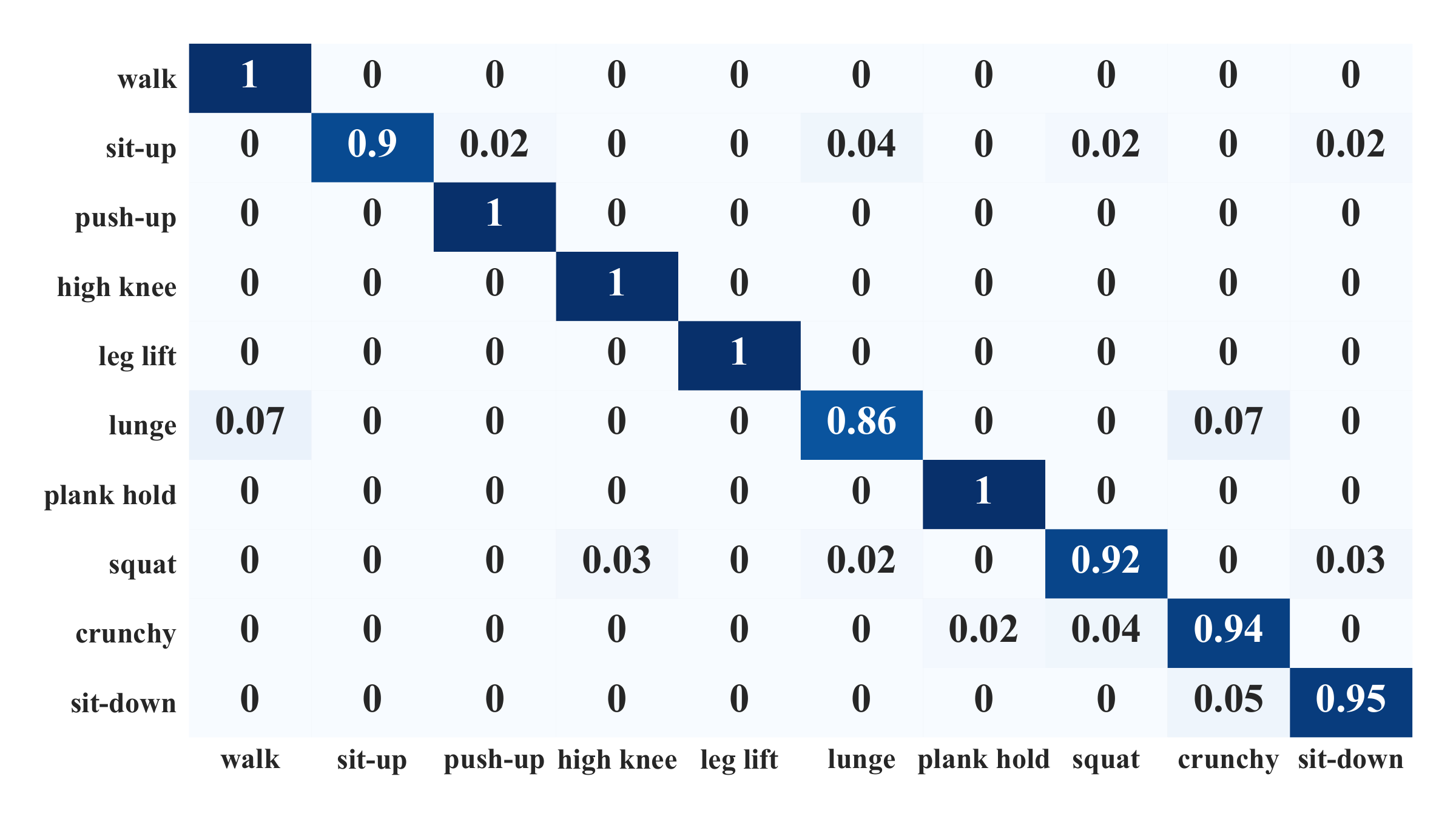}
	    \vspace{-0.25in}
	    \caption{HAR results.}
	    \label{fig:effectiveness-har}
    \end{subfigure}
    \begin{subfigure}[b]{.49\linewidth}
        \centering
        \includegraphics[width=\linewidth]{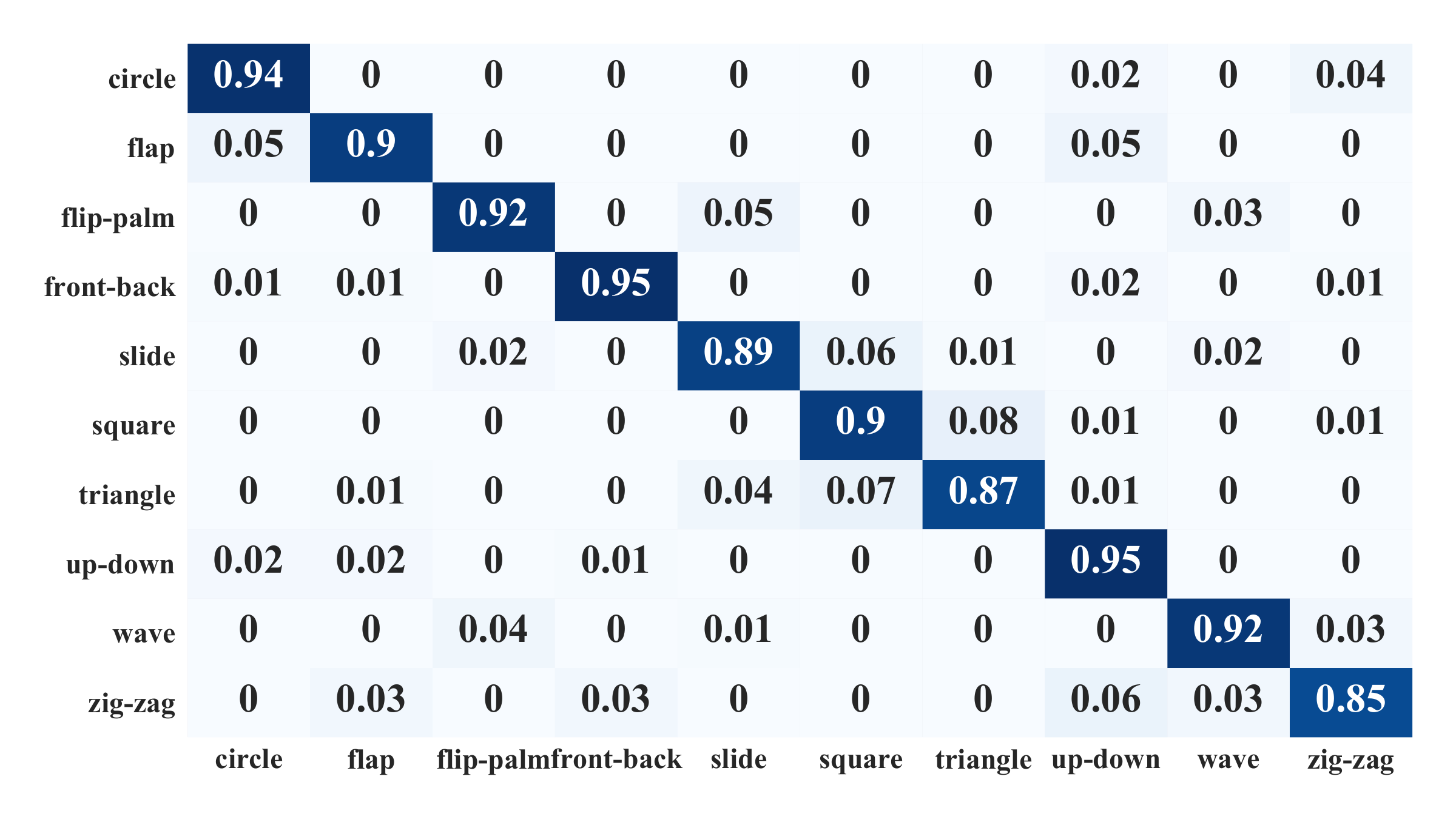}
        \vspace{-0.25in}
        \caption{HGR results.}
        \label{fig:effectiveness-hgr}
    \end{subfigure}
	\vspace{0.05in}
	\caption{Confusion matrices of the classification results of ten human activities (HAR) and ten hand gestures (HGR).}
        \vspace{-0.1in}
	\label{fig:effectiveness-classification}
\end{figure*}


\subsection{Dataset}
\label{subsec:eval-dataset}

We recruit ten participants\footnote{This work has been approved by the IRB board of our institution.} to collect three independent datasets for three different sensing tasks: respiration monitoring dataset $\mathcal{D}_{RM}$, human activity recognition dataset $\mathcal{D}_{HAR}$, and hand gesture recognition dataset $\mathcal{D}_{HGR}$. The details are illustrated below. 

\begin{itemize}[leftmargin=10pt, itemsep=1pt, parsep=1pt]
    \item $\mathcal{D}_{RM}$: We ask the participants to breathe naturally in three common conditions: lying, sitting, and standing. For each participant, we record the harvested voltage signal in one minute and repeat this process for ten times ($3\times 10\times 10=300$ minutes respiration data in total). To obtain the ground truth, each participant wears a commercial wearable device FLOWTIME headband~\cite{flowtime} during the data collection.
    \item $\mathcal{D}_{HAR}$: We ask the participants to perform ten different exercise activities including walk, sit-up, push-up, high knee, leg lift, lunge, plank hold, squat, crunchy, and sit down. Each participant is asked to perform $10$ groups of each activity with five repetitions ($10\times 10\times 10\times 5=5000$ samples in total). In the evaluation, $80\%$ of the dataset is randomly selected to train the activity recognition classifier while the rest of $20\%$ data is used for testing.
    \item $\mathcal{D}_{HGR}$: We ask the participants to perform ten different hand gestures including slide, front-back, up-down, zig-zag, wave, circle, triangle, square, flap, and flip-palm. Each participant is asked to perform $10$ groups of each gesture with five repetitions of both left-hand and right-hand ($10\times 10\times 10\times 5 \times 2=10000$ samples in total). Similarly, we use $80\%$ data samples to train the CNN neural network of the hand gesture classifier and evaluate the performance with the rest of $20\%$ data samples.
\end{itemize}

\subsection{Overall Effectiveness}
\label{subsec:eval-overall}

\paragraph{Metrics.} For respiration monitoring, we obtain the respiratory rate with the standard deviation (STD) of each participant and compare the respiratory rate produced by \sysname with the recorded ground truth. We use the classification accuracy and confusion matrix as metrics for human activity and hand gesture recognition.
\looseness=-1

\paragraph{Effectiveness of RM.} \autoref{fig:effectiveness-rm-lying}, \autoref{fig:effectiveness-rm-sitting}, and \autoref{fig:effectiveness-rm-standing} show the average respiratory rates of the ten participants from \sysname and their corresponding ground truths when the participants are lying, sitting, and standing, respectively. Among all the ten participants, the calculated respiratory rate presents a maximum error of $4.6\%$, $6.4\%$, and $13.2\%$ in three situations. The detection accuracy when the participant is lying (accuracy $95.4\%$) and sitting ($93.6\%$) is higher than that of standing (accuracy $86.8\%$). This is because when the participant is standing, the rise-and-fall of both the front chest and the back causes diffraction effects, making respiration monitoring more challenging.
In addition, we find the average respiratory rate provided by \sysname is close to the ground truth but \sysname's calculation presents a larger standard deviation (averagely STD $1.2\%$, ground truth STD: $0.59\%$) which means \sysname performs better when monitoring the respiration for a long-time duration.

\paragraph{Effectiveness of HAR.} \autoref{fig:effectiveness-har} shows the classification results (confusion matrix) of the trained CNN-based HAR classifier. The overall accuracy in recognizing the ten involved human activities is $95.7\%$. From the result, we find that \sysname can accurately recognize activities like walk, push-up, and high knee ($100\%$ accuracy), but performs the worst in recognizing activities like sit-up (accuracy $90\%$) and lunge (accuracy $86\%$) because these activities requires the person to perform similar actions, \ie, bending knees to squat, which results in close patterns of the harvested voltage signals and increases the number of misclassified activities.

\begin{figure*}[]
	\centering
	\begin{subfigure}[b]{.325\linewidth}
	    \centering
	    \includegraphics[width=\linewidth]{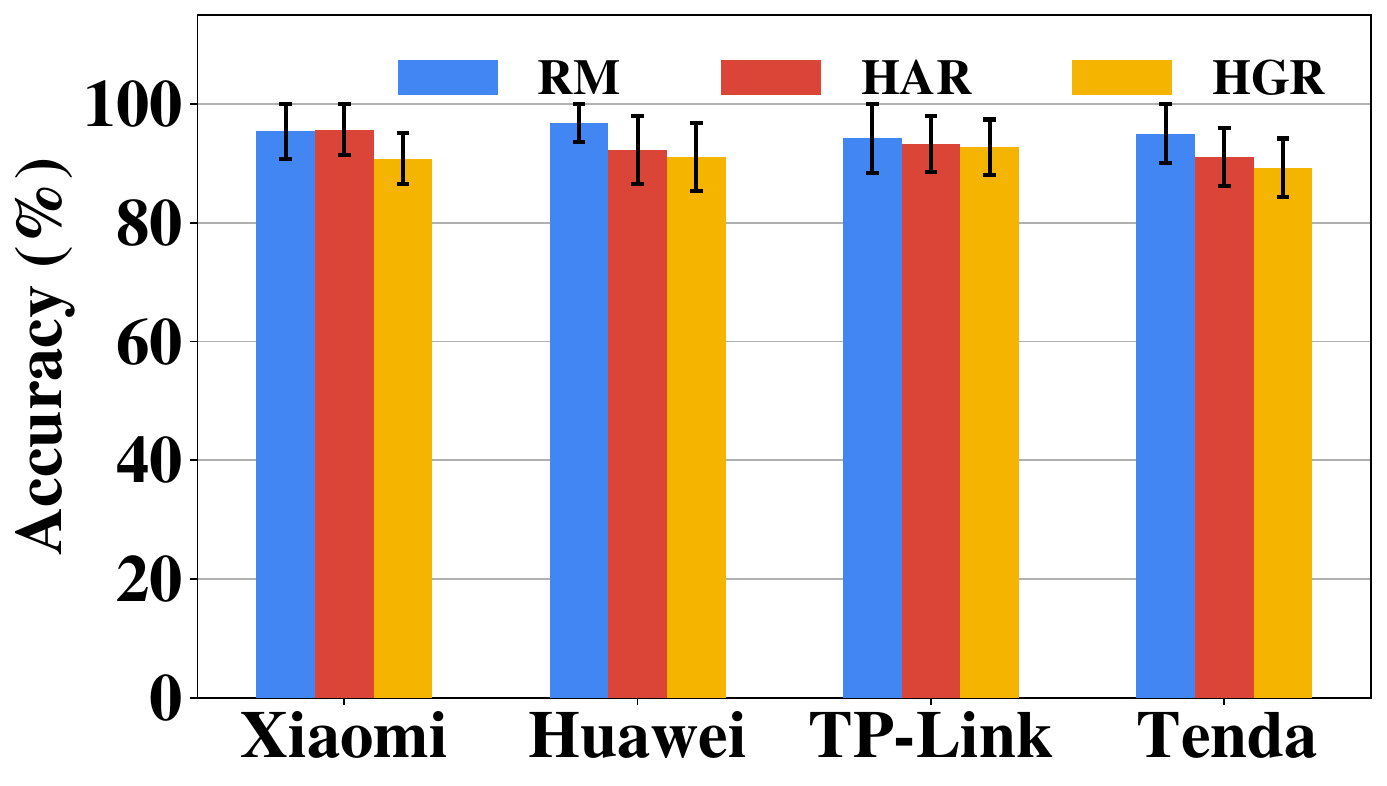}
	    \vspace{-0.2in}
	    \caption{Impact of Wi-Fi routers.}
	    \label{fig:impact-factor-router}
    \end{subfigure}
    \begin{subfigure}[b]{.325\linewidth}
        \centering
        \includegraphics[width=\linewidth]{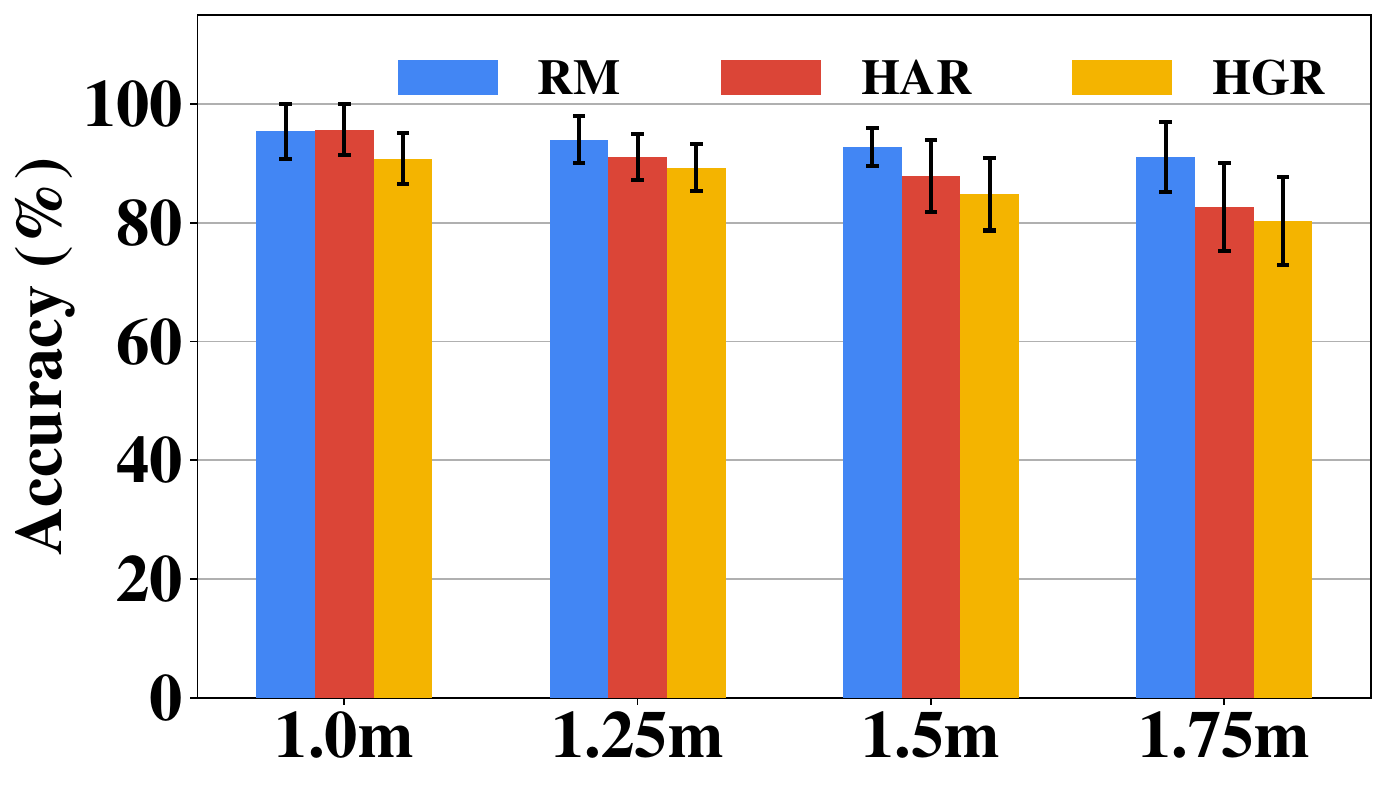}
        \vspace{-0.2in}
        \caption{Impact of sensing ranges.}
        \label{fig:impact-factor-sensing-range}
    \end{subfigure}
    \begin{subfigure}[b]{.325\linewidth}
        \centering
        \includegraphics[width=\linewidth]{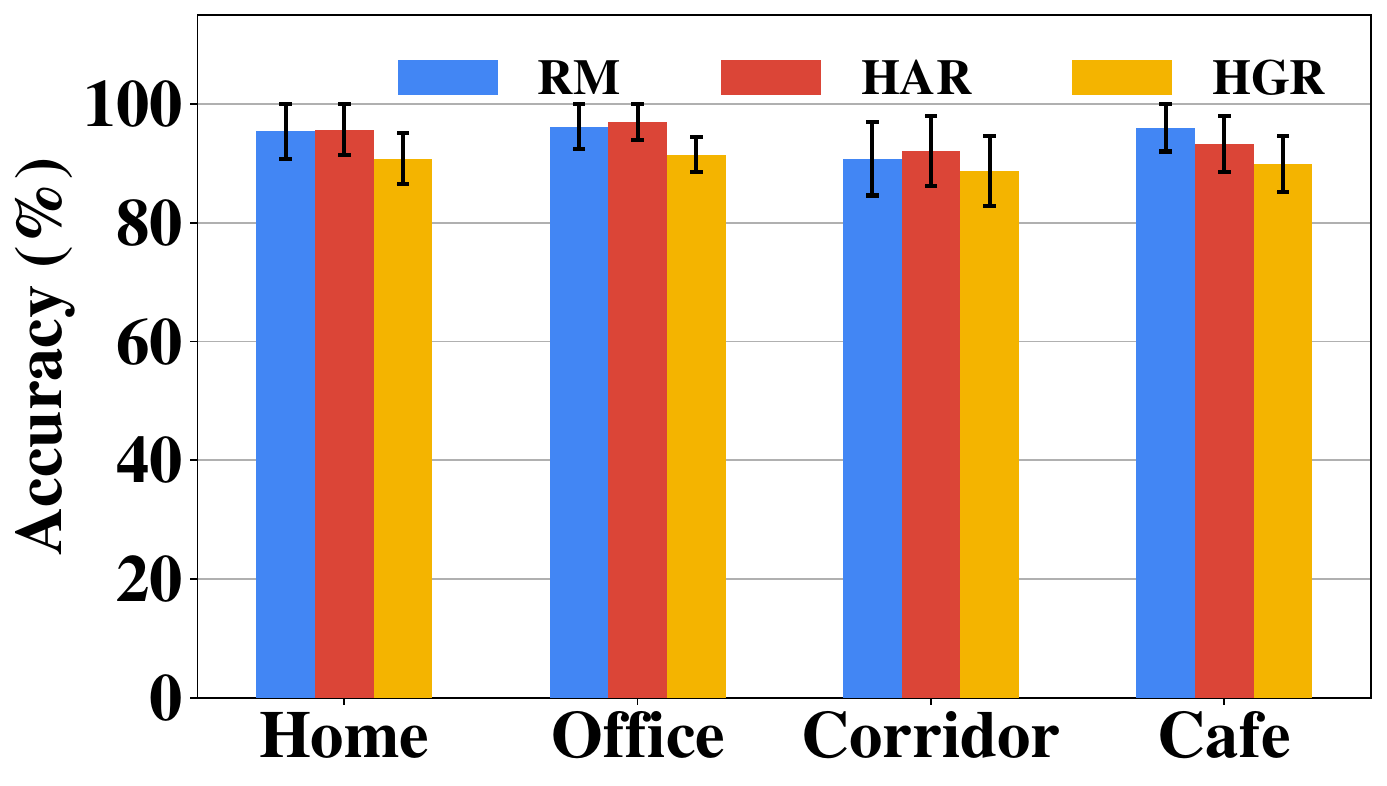}
        \vspace{-0.2in}
        \caption{Impact of environments.}
        \label{fig:impact-factor-environment}
    \end{subfigure}
	\vspace{0.05in}
	\caption{Analysis of the impact of different Wi-Fi routers, sensing ranges, and environments.}
        \vspace{-0.1in}
	\label{fig:impact-factors}
\end{figure*}

\paragraph{Effectiveness of HGR.} \autoref{fig:effectiveness-hgr} shows the classification results (confusion matrix) of the trained CNN-based HGR classifier. The overall accuracy in recognizing the ten involved hand gestures is $90.8\%$. In addition, we observe \sysname performs the best in recognizing hand gestures such as front-back and up-down (accuracy $95\%$) but presents the lowest accuracy in recognizing gestures such as zig-zag (accuracy $85\%$) and triangle (accuracy $87\%$) because gestures like zig-zag and up-down, triangle and square have similar keystrokes and tracks, which causes similar energy patterns and increases the number of misclassified gestures.

\subsection{Impact of Practical Factors}
\label{subsec:impact-factors}

\paragraph{Different Wi-Fi routers.} In \autoref{sec:evaluation}, we present the overall effectiveness of \sysname with the Xiaomi MI Router. To evaluate the impact of different commodity Wi-Fi routers and demonstrate the generalization ability of our system, we follow the same procedures in the aforementioned three sensing tasks with other three Wi-Fi routers: Huawei 4G Router 2 Pro, TP-Link TL-WR742N, and Tenda F3. \autoref{fig:impact-factor-router} shows the accuracy
in calculating the respiratory rate (lying), determining human activities, and recognizing hand gestures of different Wi-Fi routers. For the three sensing tasks, the maximum differences in accuracy rates using different Wi-Fi routers are $2.6\%$, $4.6\%$, and $3.4\%$, respectively, demonstrating the robustness of \sysname for different Wi-Fi routers.
\looseness=-1

\paragraph{Different sensing ranges.} As mentioned in \autoref{subsec:rfeh}, the harvested RF energy depends on the distance (\aka sensing range) of $Tx$ and $Rx$.
To investigate the impact of different sensing ranges, we adjust the sensing range from \SI{1.0}{\meter} (default setting) to \SI{1.25}{\meter}, \SI{1.5}{\meter}, \SI{1.75}{\meter}, and repeat the similar experiment processes. \autoref{fig:impact-factor-sensing-range} shows the classification accuracy of \sysname in different sensing tasks at different sensing ranges. We find the accuracy rates of the three sensing tasks decrease approximately by $4.3\%$, $13.1\%$, and $10.5\%$ as the sensing range changes from \SI{1.0}{\meter} to \SI{1.75}{\meter}. This is because as the sensing range increases, the FFZ becomes larger and sensitive to the surrounding noise (\ie, other transmitter sources, movements) that deteriorates the performance of \sysname. Moreover, the harvested energy decreases due to the transmission attenuation of RF signal, leading to a low signal-to-noise ratio. For example, when the Wi-Fi router and the RF energy harvesting device are separated by \SI{1.2}{\meter}, nearly $90\%$ energy is lost in the air~\cite{aboueidah2017characterization}.
Despite the sensing range in our current settings being limited (\eg, $<$\SI{1.75}{\meter}), we present \sysname as a proof of concept that we can harness RF energy harvesting for wireless sensing. In practice, the sensing distance can be extended by implementing a more powerful energy harvester with high-gain antennas and more RF-DC converters. For instance, works such as~\cite{talla2017battery} can harvest RF energy at a maximum distance of \SI{9.75}{\meter}, and we will focus on extending the sensing distance of \sysname to make it more robust in our future work.

\paragraph{Different environments.} We further conduct the same experiments in the other three common indoor scenarios (office, corridor, and cafe) to investigate the impact of environments. \autoref{fig:impact-factor-environment} presents the classification accuracy of the three sensing tasks in different environments. In different environments, \sysname shows overall accuracy rates of $94.6\%$, $94.5\%$, and $90.2\%$ for respiration monitoring, activity recognition, and gesture recognition, respectively. We notice that the maximum accuracy difference in the four environments is $5.4\%$, $4.9\%$, and $2.8\%$ only. This is because the small signal variations induced by the environment noise (\eg, other people's movements) can be removed by the S-G filter as has been proved in previous studies~\cite{zhang2019towards, zhang2021fresnel}. The results demonstrate the robustness of \sysname in different environments.

\begin{figure}[t]
    \centering
    \includegraphics[width=\linewidth]{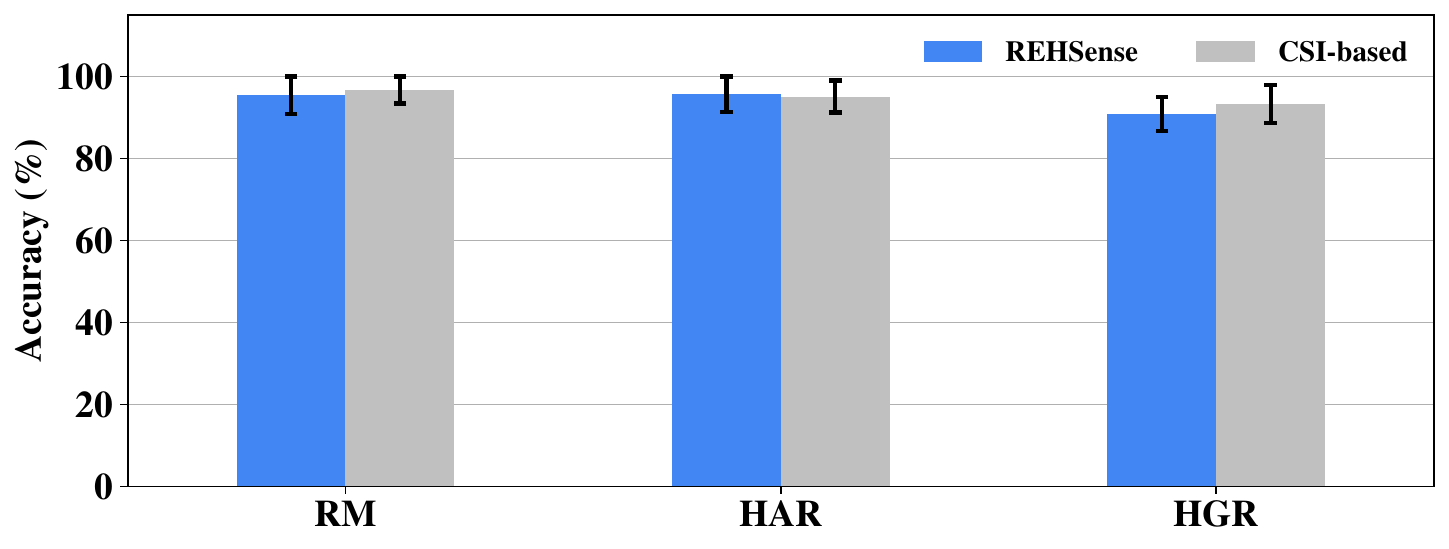}
    \caption{Comparison with the CSI-based sensing system.}
    \vspace{-0.15in}
    \label{fig:comparison-csi}
\end{figure}

\subsection{Comparison with Wi-Fi CSI-based Sensing}
\label{subsec:eval-comparison}

\paragraph{Sensing accuracy.} We now compare the performance of \sysname with the state-of-the-art Wi-Fi CSI-based approaches in the same settings. For the hardware, we install the commonly used the Linux 802.11n CSI tool~\cite{halperin2011tool} on a Lenovo T400 laptop to extract CSI. Then, we implement the methods proposed by Zhang et al.~\cite{zhang2019towards}, Gao et al.~\cite{gao2022towards}, and Zhang et al.~\cite{zhang2018from} as the baseline of activity recognition, gesture recognition, and respiratory monitoring, respectively. \autoref{fig:comparison-csi} 
compares the accuracy of \sysname with the corresponding CSI-based system in the three sensing tasks.
Specifically, \sysname and the CSI-based sensing system separately achieve accuracy of $95.4\%$ vs. $96.7\%$ in respiration monitoring, $95.7\%$ vs. $95.1\%$ in human activity recognition, and $90.8\%$ vs. $93.3\%$ in hand gesture recognition.
The result shows \sysname achieves comparable accuracy with the CSI-based systems.
\looseness=-1

\paragraph{Energy efficiency.} To demonstrate the energy-efficient of our system, we use the Monsoon Power Monitor to measure the energy consumption of each hardware component. \sysname utilizes a passive RF energy harvesting board to capture Wi-Fi signal, which does not consume any energy. The Arduino Nano (with a MicroSD card) consumes $11.3-$\SI{12.6}{\milli\watt} for data collection. By contrast, an Intel $5300$ NIC card in the CSI-based sensing system consumes $820-$\SI{940}{\milli\watt}~\cite{halperin2010demystifying, jang2011snooze} to receive Wi-Fi packets. Hence, \sysname reduces around $98.7\%$ energy consumption in the hardware. Moreover, the harvested RF energy is converted to DC voltages that can be used for charging sensors and batteries. For instance, assume an IoT device (\ie, LED lamp) is powered by a \SI{9}{\volt}, \SI{550}{\milli\ampere\hour} rechargeable battery that can support a power-intensive component like Intel $5300$ NIC for only $5.3-6.0$ hours, whereas the battery can support the low-power \sysname for around $400$ hours. In addition, \sysname can harvest approximate \SI{4.5}{\milli\watt} RF energy at the distance of \SI{1}{\meter} for charging the battery so that it can prolong the usage and lifestyle of this IoT device.

\subsection{Integrating \sysname into IoT Devices}
\label{subsec:integrate-iot-device}

\begin{figure}[t]
	\centering
	\begin{subfigure}[b]{.49\linewidth}
	    \centering
	    \includegraphics[width=\linewidth]{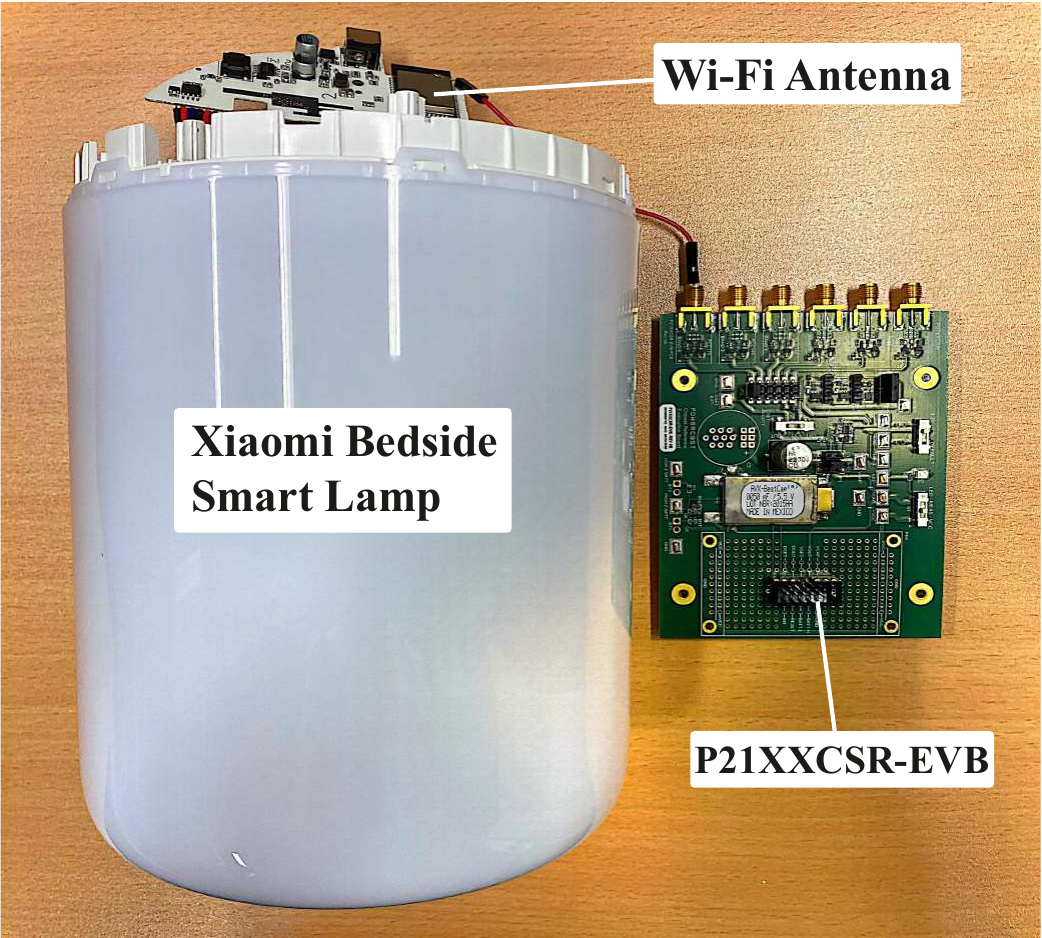}
	    \caption{w/ Xiaomi bedside lamp.}
	    \label{fig:integrate-xiaomi}
    \end{subfigure}
    \begin{subfigure}[b]{.49\linewidth}
        \centering
        \includegraphics[width=\linewidth]{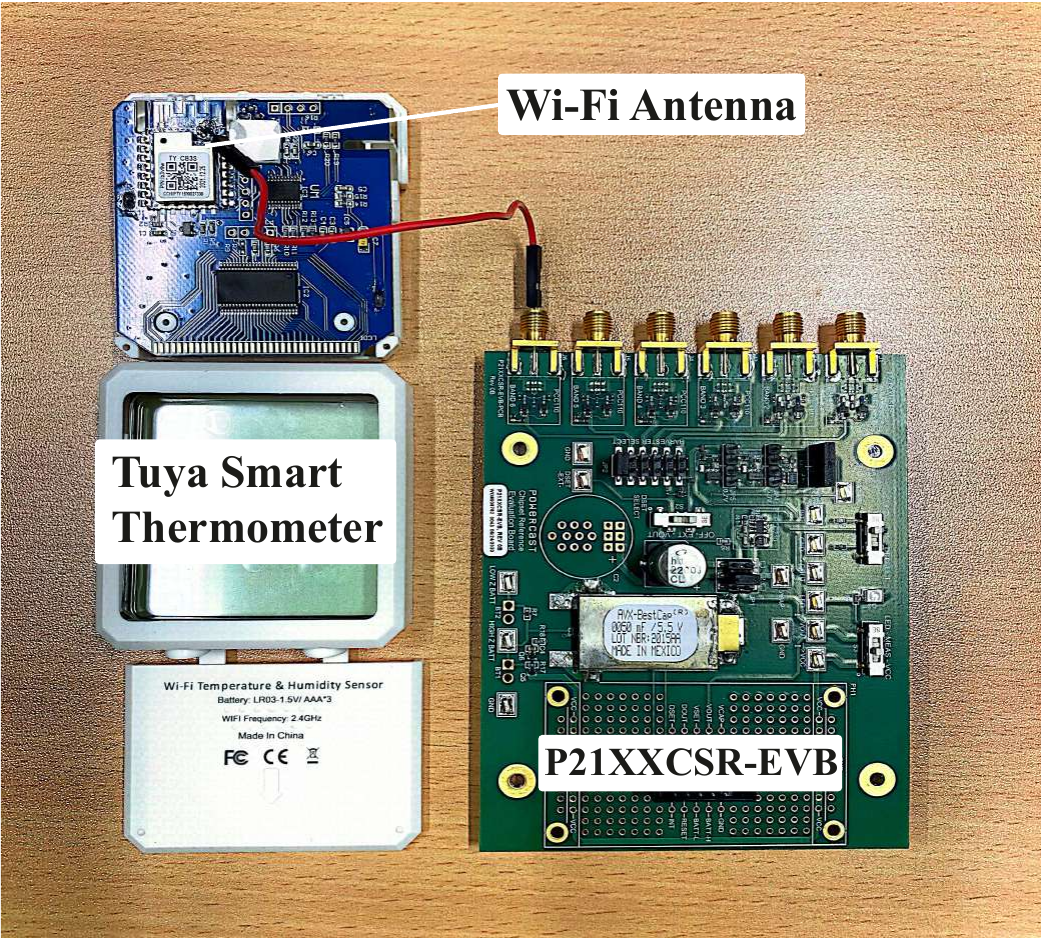}
        \caption{w/ Tuya thermometer.}
        \label{fig:integrate-tuya}
    \end{subfigure}
	\vspace{0.05in}
	\caption{Integrate \sysname into IoT devices.}
        \vspace{-0.2in}
	\label{fig:integrate-iot-setup}
\end{figure}

\begin{figure}[t]
    \centering
    \includegraphics[width=\linewidth]{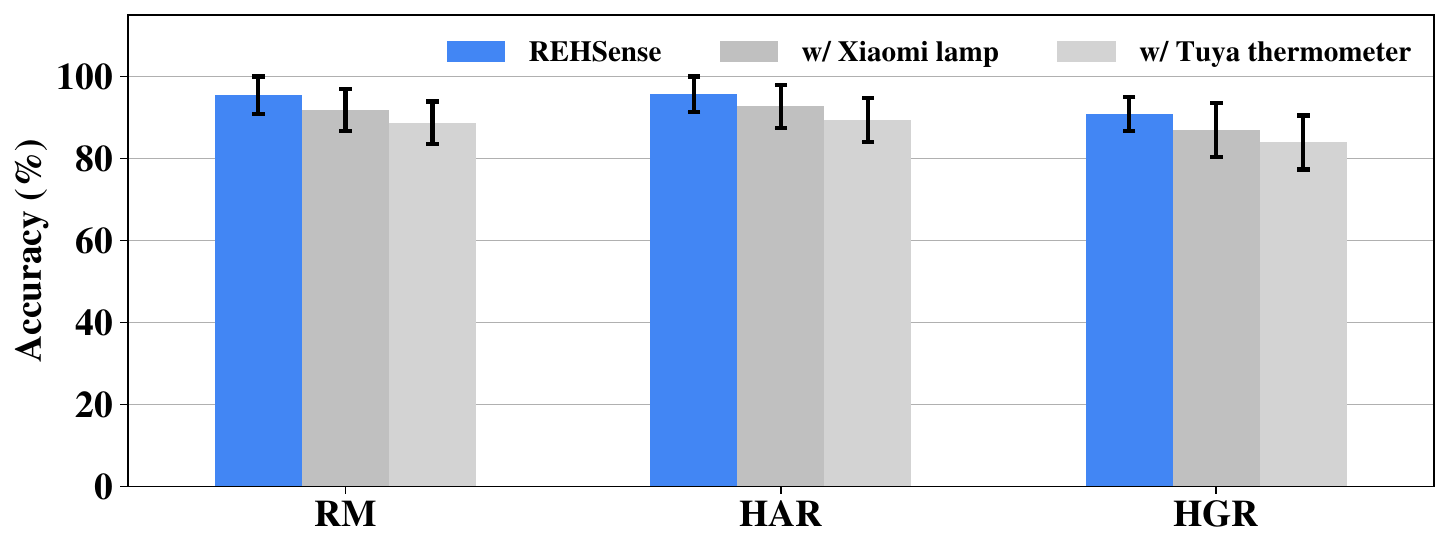}
    \caption{Effectiveness of integrating with IoT devices.}
    \label{fig:integrate-iot-result}
    \vspace{-0.2in}
\end{figure}

In this experiment, we present examples of integrating \sysname into commodity IoT devices to demonstrate the feasibility and usability of \sysname. \autoref{fig:integrate-iot-setup} shows that we integrate the \sysname into a Xiaomi bedside smart lamp and Tuya smart thermometer by connecting its on-chip Wi-Fi antenna to the RF energy harvesting board. Then, we use these two modified IoT devices as the receiver to evaluate the performance of \sysname in the three wireless sensing tasks. \autoref{fig:integrate-iot-result} presents the results of using the Xiaomi bedside smart lamp and Tuya smart thermometer as the wireless receiver (with the RF energy harvesting board), as well as the overall effectiveness we have shown in \autoref{subsec:eval-overall}. We can see that the \sysname-equipped Xiaomi bedside smart lamp shows accuracy rates of $91.8\%$, $92.7\%$, and $87.0\%$ in respiration monitoring, human activity recognition, and hand gesture recognition, respectively. The modified Tuya smart thermometer shows $88.7\%$, $89.4\%$, and $83.9\%$ accuracy in the three sensing tasks. Compared to the original \sysname prototype, the accuracy rates of the three sensing tasks drop approximately by $6\%$-$7\%$ as the inner antennas of commodity IoT devices are not powerful as an external antenna used in \sysname. 
Nevertheless, experiment results still show high feasibility and usability of \sysname in daily IoT devices. 



\begin{figure}[t]
    \centering
    \includegraphics[width=.95\linewidth]{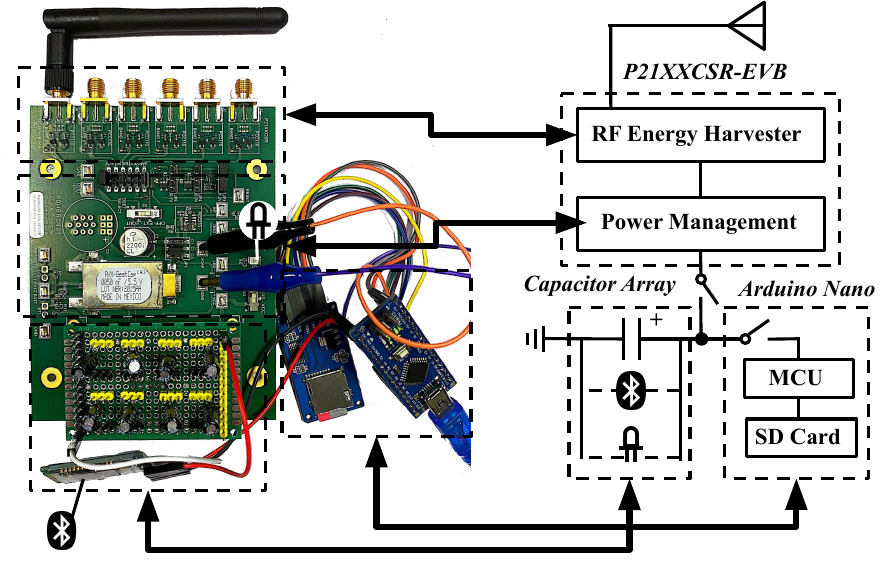}
    \caption{A prototype of the battery-free sensing system.}
    \vspace{-0.1in}
    \label{fig:batteryfree-prototype-system}
\end{figure}
\begin{figure}[t]
    \centering
    \includegraphics[width=.95\linewidth]{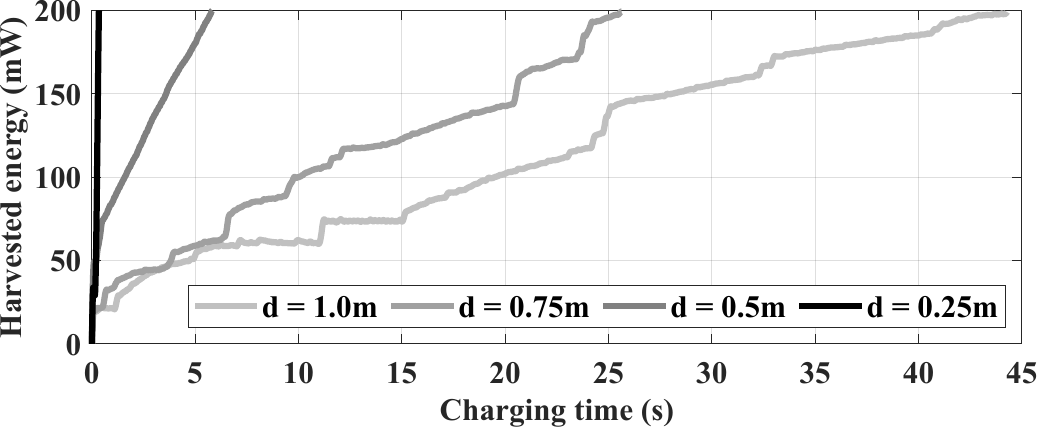}
    \caption{The harvested energy in the capacitor array vs. charging time at different distances. }
    \vspace{-0.1in}
    \label{fig:charging-capacitor-array}
\end{figure}

\subsection{Potential of Battery-free Sensing}
\label{subsec:eval-batteryfreesensing}

Since \sysname has shown promising sensing performance and low power consumption, we then implement a prototype system to demonstrate how \sysname pushes human sensing from power-hungry sensing to battery-free sensing. \autoref{fig:batteryfree-prototype-system} presents the prototype system as well as its circuit diagram, which consists of three components: an RF energy harvesting board (P21XXCSR-EVB), a storage capacitor array made by eight parallel \SI{330}{\uF} capacitors, and an Arduino Nano MCU with a MicroSD card. Specifically, the capacitor array is designed for storing harvested energy and then using the storage to support electrical components~\cite{talla2017battery, saffari2019battery}, while the Arduino Nano and MicroSD card record the DC voltages for different sensing tasks. Below, we provide two application examples of the prototype system to illustrate how to achieve battery-free sensing based on our prototype of \sysname.

\paragraph{Application 1: Sensing and Powering.} In \autoref{subsec:integrate-iot-device}, we present \sysname can be integrated into commodity IoT devices and achieve promising sensing accuracy. It also inspires us to think: \textit{can we achieve battery-free sensing by using the harvested energy for sensing and powering simultaneously?} With this question, we further optimize the above prototype by adjusting the sensing range to obtain a high harvested voltage and then explore the possibility that the captured RF energy can be used for lighting a LED and sensing without any external power sources (\ie, battery, USB powerline). Specifically,  we utilize a B\&K Precision 2190E 2-channel digital oscilloscope to monitor the changes in the harvested RF energy when the LED keeps lighting (harvested voltage $\ge$ \SI{1.6}{\volt}). We find the oscilloscope shows distinctive patterns of different hand gestures at a sensing range of \SI{10}{\centi\meter} while the harvested RF energy can also power the surface-mount LED module simultaneously. Even with a limited sensing range, this example shows that \sysname has the potential to achieve wireless sensing and power small IoT devices such as LED lights simultaneously, enabling various self-powered devices with wireless sensing capability.
\looseness=-1

\paragraph{Application 2: Sensing and Communication.} Since we have demonstrated that the harvested energy can be used to support electrical components like LED modules, another question arises: \textit{can we use the harvested RF energy to support communication so that we can transmit the harvested voltage data to the remote devices for different sensing tasks?} As the RF energy harvester can only store limited energy, we attach a capacitor array to store more harvested RF energy to achieve the application of supporting power-intensive components such as a Bluetooth module. Specifically, we use the Bluetooth module HC-05 that is running over a \SI{3.3}{\volt} voltage with \SI{200}{\milli\watt} power to support the communication, and it takes extra time to charge the capacitor array for storing enough power to achieve one data transmission process. We further explore the charging time with stored harvested RF energy at different sensing ranges from \SI{1.0}{\meter} to \SI{0.75}{\meter}, \SI{0.5}{\meter}, \SI{0.25}{\meter} and record the harvested RF energy in charging the external capacitor array so that it has enough energy to power the Bluetooth module. \autoref{fig:charging-capacitor-array} presents the time-voltage curves with different sensing ranges when the harvested RF energy is monitored for sensing and being used for charging the capacitor array. We find as the sensing range decreases, the charging time also decreases since much more Line-of-Sight RF energy is harvested by our devices. For instance, it takes \SI{44.1}{\second}, \SI{25.6}{\second}, \SI{6.0}{\second}, and \SI{0.4}{\second} to store enough energy to support one Bluetooth communication with sensing ranges of \SI{1.0}{\meter}, \SI{0.75}{\meter}, \SI{0.5}{\meter}, and \SI{0.25}{\meter}, respectively. The results reflect that a trade-off between the charging time and the sensing range should be considered to make \sysname in realizing battery-free sensing and communications.

The two experiments above demonstrate the potential of using RF energy harvesting to achieve ubiquitous battery-free sensing and corresponding applications, and \sysname makes the first step towards this goal.

\subsection{Comparison with Other Sensing Techniques}

To further illustrate the advantages of the proposed RF energy harvesting-based wireless sensing technique, we compare \sysname with four types of well-known sensing techniques: vision-based sensing~\cite{zhao2018through, li2019making} (\ie, camera-based), IMU-based sensing~\cite{kwapisz2011activity, dehzangi2018imu} (\ie, accelerometer, gyroscope), kinetic energy harvesting (KEH) based sensing~\cite{khalifa2017harke, xu2017keh}, and RFID-based backscatter sensing systems~\cite{li2016deep, ding2017platform}. \autoref{tab:comparison-other-techniques} shows the comparison results from four metrics: (1) intrusive or non-intrusive sensing. (2) Coarse-grained or fine-grained recognition. (3) Active or passive sensing. (4) power-hunger or energy efficient.
\looseness=-1

Compared with vision-based sensing systems~\cite{zhao2018through, li2019making} that use cameras to capture images and videos for human sensing, \sysname shows high efficiency in energy consumption since the RF energy harvester has no power consumption to monitor the power traces of the transmitting RF signals. IMU-based sensing systems~\cite{kwapisz2011activity, dehzangi2018imu} are popular since most commodity mobile devices have equipped with IMU sensors (\ie, accelerometer, gyroscope, and magnetometer). However, these IMU-based systems require the user to hold their smartphones or wear a smartwatch to intrusively obtain the data traces resulting from different human activities or hand gestures. To move a step to the concept of battery-free sensing, recent works~\cite{xu2017keh, khalifa2017harke} proposed new techniques named kinetic energy harvesting (KEH) that convert kinetic energy to current traces, and then use these traces for human sensing. Nevertheless, KEH-based sensing systems also require the user to wear specific sensors and can only recognize coarse-grained human activities~\cite{khalifa2017harke}. Furthermore, RFID-based sensing systems~\cite{li2016deep, ding2017platform} seem to share the same principles of \sysname as passive RFID tags also integrated with RF energy harvesting circuits to scavenge the RF energy sent by the RFID reader. However, RFID-based sensing systems need a professional RFID reader or USRP to send RF signals at approximately \SI{915}{\mega\hertz} so that they cannot be integrated into a smart home, whereas \sysname harvests RF energy from the Wi-Fi signals (\eg, \SI{2.4}{\giga\hertz}) radiated from Wi-Fi routers that are widely deployed in most indoor scenarios and also demonstrate the feasibility of being totally battery-free. Therefore, \sysname is much more practical and energy-efficient than RFID-based backscatter sensing systems.

\begin{table}[t]
\centering
\caption{Comparison with other four types of wearable/wireless sensing techniques. \ding{108} for ``Yes'', \ding{109} for ``No'', \ding{119} for ``Partially''.}
\scriptsize
\setlength{\tabcolsep}{4pt}
\label{tab:comparison-other-techniques}
\begin{tabular}{ccccc}
\toprule
\textbf{Sensing system} & \textbf{Non-intrusive?} & \textbf{Fine-grained?} & \textbf{Passive?} & \textbf{Energy-efficient?} \\ \midrule
\multicolumn{1}{c}{Vision-based~\cite{zhao2018through, li2019making}} & \ding{108} & \ding{108} & \ding{109} & \ding{109} \\ \midrule
\multicolumn{1}{c}{IMU-based~\cite{kwapisz2011activity,dehzangi2018imu}} & \ding{109} & \ding{108} & \ding{109} & \ding{109} \\ \midrule
\multicolumn{1}{c}{KEH-based~\cite{khalifa2017harke, xu2017keh}} & \ding{109} & \ding{109} & \ding{108} & \ding{108} \\ \midrule
\multicolumn{1}{c}{RFID-based~\cite{li2016deep, ding2017platform}} & \ding{108} & \ding{108} & \ding{108} & \ding{119} \\ \midrule
\multicolumn{1}{c}{\textbf{\sysname}} & \ding{108} & \ding{108} & \ding{108} & \ding{108} \\ \bottomrule
\end{tabular}%
\vspace{-0.2in}
\end{table}

\section{Related Works}
\label{sec:related}

\paragraph{Wearable sensing.} Wearable sensing techniques have been widely studied over the past few decades, which exploit sensor data collected from wearable devices to enable various sensing applications~\cite{duan2024f2key, ni2024sensor, meteriz2021sia, meteriz2022keylogging, meteriz2022acoustictype, mohaisen2023understanding, lyu2024earda, alghamdi2024xr, sun2024earpass, xu2017sensor}. 
For instance, IMU-based systems utilize IMU sensors (\ie, accelerometer, gyroscope, and magnetometer) equipped in mobile devices (\ie, smartphone, smartwatch) to infer human activities~\cite{kwapisz2011activity}, translate gesture-based sign language~\cite{li2022gasla}, and realizes automatic key generation for on-body communication~\cite{xu2016walkie}.
Cao \etal utilize the built-in accelerometer and magnetometer of smart terminals to extract information representing user identity, achieving convenient and efficient authentication for heterogeneous IoT devices~\cite{9928387, 9546516}.
Instead of requiring the user to wear sensor-based devices, \sysname introduces a novel non-intrusive sensing method, RF energy harvesting, and achieves fine-grained wireless sensing tasks with high effectiveness as well as low power consumption~\cite{ni2024rehsense}.

\paragraph{Wireless sensing.} Wireless sensing has been widely studied because of its promising performance and contact-free manner~\cite{ni2021simple, ni2023eavesdropping, ni2023exploiting, ni2023recovering, ni2023uncovering, guo2018real, ni2023xporter, zhu2018indoor, ni2021explore, sun2023flora, sun2024flora+, sun2024rf, sun2023demo, han2024mmsign, song2023emma, song2022mobilekey, chen2022swipepass, wu2024xsolar, cao2024security, hu2023earsonar, liu2021wavoice, liu2023wavoid}.
Most existing wireless sensing works~\cite{liu2015contactless, zeng2018fullbreathe,wang2017device, he2015wig, wang2017phasebeat} focus on extracting Wi-Fi CSI as the channel measurement because of its granularity and compatibility with COTS devices. 
Recent CSI-based sensing studies have formulated the theoretical description of the Fresnel diffraction sensing model and the results indicate good identification performance and generalization ability in respiration monitoring~\cite{niu2021understanding, zhang2018from}, activity recognition~\cite{zhang2019towards}, and gesture recognition~\cite{gao2022towards}. However, almost all CSI-based sensing systems need to modify the transceivers and use power-intensive Wi-Fi NICs to capture wireless channel information. 
Furthermore, WiKI-Eye~\cite{hu2023password} shows the beamforming feedback information (BFI) could be exploited to infer user credentials such as smartphone passwords.
Compared with these works, \sysname has demonstrated competitive human sensing performance without modifying the COTS devices. In addition, our system requires much less energy consumption than a conventional CSI-based system, and the harvested RF energy can be reused to power other electronic components such as LED lights and Bluetooth modules.



\section{Conclusion}
\label{sec:conclusion}

In this paper, we present \sysname, a novel wireless sensing system based on RF energy harvesting. We address two major limitations of existing Wi-Fi based wireless sensing systems by using RF energy for the dual purpose of energy harvesting and sensing. We design and implement a prototype of \sysname and comprehensively evaluate its performance in three common human sensing tasks: respiration monitoring, human activity recognition, and hand gesture recognition. Experiments show that \sysname achieves high accuracy in these sensing tasks for different routers and environments. Comparison with a traditional Wi-Fi based sensing system shows that \sysname achieves comparable accuracy and good adaptation ability while reducing energy consumption significantly. Moreover, we demonstrate the feasibility of \sysname by integrating it into commodity IoT devices and implement a prototype to demonstrate the potential of battery-free sensing.
We envision the wide deployment of \sysname when future smart IoT devices are equipped with RF energy harvesting techniques.
\looseness=-1

\begin{acks}
We sincerely thank Prof. Haiming Jin for shepherding our paper and the anonymous reviewers for their constructive comments.
This work was supported by Hong Kong RGC (Project No. CityU 21201420/11201422), the Innovation and Technology Commission of Hong Kong (Project No. PRP/037/23FX and MHP/072/23), NSF of Shandong Province (Project No. ZR2021LZH010), and NSF of Guangdong Province (Project No. 2414050001974).
Any opinions, findings, and conclusions in this paper are those of the authors and not necessarily of the supported organizations.
\end{acks}


\balance
\bibliographystyle{unsrt}
\bibliography{sample-base}










\end{document}